\documentclass[12pt]{report}
\usepackage{graphics}
\include{graphics}
\begin{document}
\topmargin-1cm
\baselineskip24pt 
\begin{titlepage}
\null
\begin{center}
{\bf \normalsize QUANTUM HAMILTON - JACOBI STUDY OF WAVE FUNCTIONS AND
ENERGY SPECTRUM OF SOLVABLE AND QUASI - EXACTLY SOLVABLE MODELS} \\
\vfill
{\normalsize A thesis submitted in partial fulfilment of the requirements\\
 for the award of the degree of} \\
\vfill
{\normalsize
{\bf DOCTOR OF PHILOSOPHY}\\
{\it in}\\
{\bf PHYSICS}\\
{\it by}\\
{\bf K. G. GEOJO}}\\
%\vskip0.5cm
\vfill
\includegraphics*{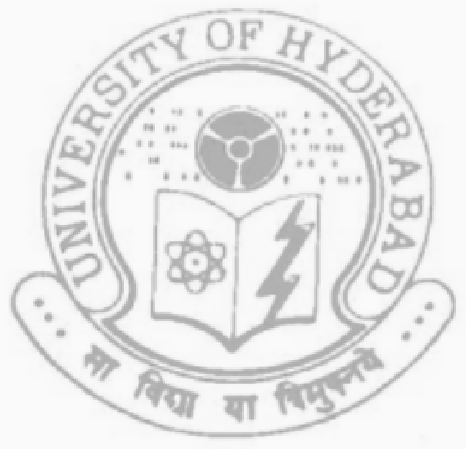}
%\input epsf
%\psfxsize5.0cm\psfbox{uhsealc1.eps}\\
%\vskip0.5cm
\vfill
{\bf\small SCHOOL OF PHYSICS\\UNIVERSITY OF HYDERABAD\\ HYDERABAD - 500046,
INDIA\\ DECEMBER 2003\\}
\end{center}  
\end{titlepage}

%completed 0n 2 julu
\begin{center}
\null
\vskip2.3cm
{\bf \Large Declaration}
\end{center}
\vskip1.5cm
     I, \textbf{\textit{K. G. Geojo}}, hereby declare that the
     work reported in this dissertation  
titled,  \textbf{\textit{Quantum Hamilton - Jacobi study of wave
     functions and energy spectrum of solvable and quasi - exactly
     solvable models}}, is
entirely original and has been carried out by me,  under the supervision of
\textbf{\textit{Prof. A. K. Kapoor}}, Department of Physics, School of
Physics, University of Hyderabad.
   
     To the best of my knowledge, no part of this dissertation was
submitted for any degree of any other institute or university.\\ \\ \\ \\ \\
Place: Hyderabad \\
Date:\hfill K. G. Geojo

%completed on 2 july
\begin{center}
\null
\vskip2.3cm
{\Large \bf Certificate}
\end{center}
\vskip1.5cm
    This is to certify that the report entitled
\textbf{\textit{Quantum Hamilton - Jacobi study of wave functions and
    energy spectrum of solvable and quasi - exactly solvable models}}, being
submitted by \textbf{\textit{K. G. Geojo}}, in partial fulfillment of the
requirements for the award of \textbf{\textit{Doctor of Philosophy
in Physics}} by \textbf{\textit{University of Hyderabad}}, is a
bonafide work carried out at the \textbf{\textit{University of
Hyderabad}} under my supervision. The matter embodied in this report
has not been submitted to any other institute or university for the
award of any degree.\\ \\ \\ \\ \\
\begin{minipage}[b]{4.6cm}
\baselineskip 18pt
{\bf Dean},\\
School of Physics,\\
University of Hyderabad,\\
Hyderabad - 500046.
\end{minipage}
\hfill
\begin{minipage}[b]{4.6cm}
\baselineskip 18pt
\flushright{{\bf Prof. A. K. Kapoor},\\
School of Physics,\\
University of Hyderabad,\\
Hyderabad - 500046.}
\end{minipage}

\tableofcontents
\chapter{
INTRODUCTION}

In this thesis we present an alternative approach to the study of
exactly solvable and quasi-exactly solvable (QES) problems in
quantum mechanics.  This approach, known as quantum
Hamilton-Jacobi (QHJ) approach, [1] has been found to be an
elegant and simple method to determine the energy spectrum of
exactly solvable models in quantum mechanics.  The advantage of
this method is that it is possible to determine the energy
eigen-values without having to solve for the eigen-functions.  In
this formalism, a quantum analog of classical action angle
variables [2] is introduced.  An exact quantization condition is
formulated as a contour integral, representing the quantum action
variable, in the complex plane.  This exact quantization condition
has been utilized for determining the energy eigen-values for one
dimensional and separable systems.  The quantization condition
represents well known results on the number of nodes of the
wave-function, translated in terms of logarithmic derivative, also
called quantum momentum function (QMF).   The equation satisfied
by the QMF is a non-linear differential equation, called quantum
Hamilton-Jacobi equation leads to two solutions.  A boundary
condition --- in the limit $ \hbar \rightarrow 0 $ QMF tends to
the classical momentum
--- is used to determine physically acceptable solutions for the
QMF. The application of QHJ to eigen-values has been explored in
great detail by Bhalla et al [3,4].

In chapter 2 we  review  QHJ method and, by means of an example,
we show how eigen-values are calculated without the need to obtain
the full wave function.  Briefly, this is possible because for the
implementation of exact quantization condition one needs the
knowledge of the singularities of QMF and the residues.  The
residues are easily computed by substituting only a few terms of
the Laurent's expansion in the QHJ equation.

In chapter 3 we show how to calculate bound state wave functions
in the QHJ formalism. For this purpose,  again, one only requires
knowledge of singularities of QMF and the corresponding residues.
The technique to  calculate residues is already available from
earlier works  and these are used for obtaining the bound state
wave functions.  In this process we clarify certain assumptions
which are needed, and are found to be correct, for all the exactly
solvable models which we have studied. As a by-product of the
study of the bound state wave functions we have another way of
obtaining the energy eigen-values.  In this chapter we present
details of our calculation for harmonic oscillator, hydrogen atom,
Poschl Teller, Morse  and Eckart Potentials, while details of some
other potentials can be found in [5].

In chapter 4 we take up a  study of the QES potential models [6]
in one-dimension.  These models  have been extensively studied
using Lie algebras.  The QES models have the property that a part
of the energy spectrum and corresponding wave-functions can be
computed exactly if the potential parameters satisfy a constraint
known as the condition for quasi exact solvability.  We study
several QES models within the frame work of QHJ method. In each
case we show  that the condition for quasi exactly solvability
follows from a very simple assumption about the behavior of QMF at
infinity. Our assumption is equivalent to assuming that, after a
suitable transformation, the QMF reduces to a rational function of
the independent variable.  For all the known QES models, the
quasi-exact solvability condition can be derived in this fashion
[7].

In chapter 5 we study the wave functions of quasi-exactly solvable
models and present details of our calculation for sextic
oscillator and hyperbolic potential.  We find that obtaining
eigen-values and eigen-functions does not require any new
technique other than those given in chapter 3 for the exactly
solvable models.  However, this study reveals an interesting
property of QMF for the bound states of quasi-exactly solvable
models.  This result concerns the zeros of the bound state
wave functions in the complex plane.  In the case of exactly
solvable models, the moving poles of QMF appear only on the real
line and all such poles correspond to the nodes of the wave-
function.  The number of such poles increases with energy in
accordance with well known theorems on the  number nodes of the
wave-function.  In the case of QES models all the bound state
wave-functions, which are computable algebraically and also by our
method, have complex zeros in addition to  the real zeros
corresponding to the nodes. {\it \bf{ In fact we find that for a
given QES potential all such wave-functions have the same number
of zeros if we count all real and complex zeros.}}

In the last chapter, we give a summary of our work as well as some
directions for further investigations within the QHJ formalism.

%XXXXXXXXXXXXXXXXXXXXXXXXXXXXXXXXXXXXXXXXXXXXXXXXXXXXXXXXXXXXXXXXXXXXXXX
%XXXXXXXXXXXXXXXXXXXXXXXXXXXXXXXXXXXXXXXXXXXXXXXXXXXXXXXXXXXXXXXXXXXXXXX
%XXXXXXXXXXXXXXXXXXXXXXXXXXXXXXXXXXXXXXXXXXXXXXXXXXXXXXXXXXXXXXXXXXXXXXX
\chapter{QUANTUM HAMILTON-JACOBI FORMALISM}

In this chapter, we summarize the main results of the
Hamilton-Jacobi theory in classical mechanics and the QHJ
formalism to be used in this thesis.  In section 3 a quantization
condition is given, which is exact for one dimensional system and
separable systems in higher dimensions.  In section 4 the
connection of QHJ formalism and Schr$\ddot{o}$dinger quantum
mechanics is spelled out and in the last section of this chapter
an example of computation of eigen-values within the QHJ
formalism, and without solving for wave-functions, is given.

\section{Classical Hamilton-Jacobi Theory}

The phase-space formalism of classical mechanics gives us freedom
to introduce a pair of cannonical variables $(Q_{k},P_{k})$ which
are functions of a given starting set of variables
$(q_{k},p_{k})$.  The Hamiltonian form of equations of motion is
preserved if the transformation is cannonical in the sense of
preserving Poisson brackets.  This freedom is utilized in the
Hamilton-Jacobi theory to give a formal solution of classical
mechanical problems by making a transformation, so that the
Hamiltonian becomes constant.

In the Hamilton-Jacobi theory we look for a function
$W(q_{i},P_{i})$, which generates the desired cannonical
transformation making the Hamiltonian a constant. The
transformation equation relating the old and new cannonical
variables are
\begin{equation}
p_{i}=\frac{\partial{W}}{\partial{q_{i}}}  \ , \
Q_{i}=\frac{\partial{W}}{\partial{P_{i}}}
\end{equation}
and the requirement, that the Hamiltonian in terms of new
variables be a constant $\alpha _{1} $, gives a partial
differential equation for $W(q_{i},P_{i})$:
\begin{equation}
H\left(q_{i},\frac{\partial{W}}{\partial{q_{i}}}\right)=\alpha
_{1}.     \label{2.1a}
\end{equation}
This equation is the Hamilton-Jacobi equation.  The function
$W(q_{i},P_{i})$ is known as the Hamilton's characteristic
function. It is well known that a solution to the Hamilton-Jacobi
equation is equivalent to full solution of Euler Lagrange
equations of motion [2].

In general $W(q_{i},P_{i})$, which is a function of $q_{i}$ and
$P_{i}$, can be taken to be a function of $q_{i}$'s and constants
of motion  $ \alpha _{1},\alpha _{2},\cdots ,$  by identifying the
new momenta $P_{i}$ with the constants of motion $\alpha _{i}$,
where $\alpha _{1}$ is the total energy of the Hamiltonian.

For the purpose of finding the frequencies without solving the
equation of motion completely,  action variable $J_{i}$ is
introduced by
\begin{equation}
J_{i}=\oint {\frac{\partial{W(q_{i},\alpha _{1},\alpha _{2},\cdots
,\alpha _{n})}}{\partial{q_{i}}} }dq_{i}
\end{equation}
where the integral is over a periodic orbit.  The action variable
$J_{i}$ are functions of the constants of motion $\alpha_{i}$ and
one can eliminate $\alpha_{i}$'s in favor of the action variables
$J_{i}$.  In particular the Hamiltonian, $H=\alpha_{1}$, can now
be expressed in terms of action variable $J_{i}$
\begin{equation}
H=H\left(J_{1},J_{2},\cdots,J_{n}\right)
\end{equation}

The generalized phase-space variable conjugate to $J_{i}$ are
known as the angle variable $\omega_{i}$ and are given by the
transformation equations
\begin{equation}
\dot{\omega}_{i}=\frac{\partial{H(J_{1},J_{2},\cdots,J_{n})}}{\partial{J_{i}}}
=\nu_{i}\left(J_{1},J_{2},\cdots,J_{n}\right)
\end{equation}
where the $\nu_{i}$'s are a set of constant functions of the
action variables.  The equation has the solution
\begin{equation}
\omega_{i}=\nu_{i}t+\beta_{i}.       \label{2.1b}
\end{equation}
The constant $\nu_{i}$ are just the frequencies associated with
the periodic motion, and  $\beta_{i}$'s are constants of
integration.  This formalism  (\ref{2.1b}) then gives, the
frequencies of periodic motion.  The semi-classical
Bohr-Sommerfeld quantization rule is obtained if we require that
the action-variables are integral multiples of Planck's constant.

We will now summarize the QHJ formalism and give an exact
quantization rule and its relationship with Schr$\ddot{o}$dinger
formalism.  We will first give the QHJ equation for one
dimensional system which can be easily generalized for a separable
system in several dimensions.

%XXXXXXXXXXXXXXXXXXXXXXXXXXXXXXXXXXXXXXXXXXXXXXXXXXXXXXXXXXXXXXXXXXXXXXXXXXX
%XXXXXXXXXXXXXXXXXXXXXXXXXXXXXXXXXXXXXXXXXXXXXXXXXXXXXXXXXXXXXXXXXXXXXXXXXXX
\section{Quantum Hamilton-Jacobi Equation }

In the quantum theory,  ( with $2m=1$), one assumes the generating function
$W(x,E)$ satisfies
\begin{equation}
\frac{\hbar}{i}\frac{\partial^{2}{W(x,E)}}{\partial{x^{2}}} +
\left[\frac{\partial{W(x,E)}}{\partial{x}}\right]^{2} =E-V(x)
\label{2.2a}
\end{equation}
which will be called the quantum Hamilton-Jacobi (QHJ) equation.
The momentum function
\begin{equation}
p(x,E)=\frac{\partial{W(x,E)}}{\partial{x}}
\end{equation}
will be called the quantum momentum function (QMF).  In the limit
$\hbar\rightarrow 0$ the QHJ equation goes over to the classical
Hamilton-Jacobi equation (\ref{2.1a}).  Also the QMF tends to
classical momentum function
\begin{equation}
p(x,E) \stackrel{\hbar \rightarrow 0}\longrightarrow p_{cl}(x,E)
=\left[E-V(x)\right]^{\frac{1}{2}}  \label{2.2b}
\end{equation}
From (\ref{2.2a}) it is seen that the QMF satisfies the following
equation
\begin{equation}
p^{2}(x,E)-i\hbar p^{\prime}(x,E) -[E-V(x)]=0  \label{2.2c}
\end{equation}
This equation will also be referred to as the QHJ equation.

%XXXXXXXXXXXXXXXXXXXXXXXXXXXXXXXXXXXXXXXXXXXXXXXXXXXXXXXXXXXXXXXXXXXXXXXXXXX
%XXXXXXXXXXXXXXXXXXXXXXXXXXXXXXXXXXXXXXXXXXXXXXXXXXXXXXXXXXXXXXXXXXXXXXXXXXX
\section{Exact Quantization}
\subsection{Boundary condition}

The QHJ equation (\ref{2.2c}) is a non-linear differential equation and
will give rise to two solutions. We need to establish a boundary
condition to select physically acceptable solution.  We first
summarize the original boundary condition proposed by Leacock and
Padgett. We state the boundary condition, which will complete the
definition of the QMF $p(x,E)$ in terms of the classical momentum
function $p_{cl}(x,E)$. The classical momentum function  $
p_{cl}(x,E)$  defined by (\ref{2.2b}) is a multi-valued function
of $x$ and  is defined by the following rule:

The turning points  $x_{1}$  and  $x_{2}$ are defined by the
vanishing of  $p_{cl}(x,E)$ \ i.e   \ by $
p_{cl}(x_{1},E)=p_{cl}(x_{2},E)=0$.  The complex $x$ plane on
which $p_{cl}(x,E)$ is defined, is given a cut connecting the two
branch points,  \ i.e.,  \ a cut from $x_{1}  $  to $x_{2}$. $
p_{cl}(x,E) $ is defined as that branch of the square root, which
is positive along the bottom of the cut.

With the above definition of the classical momentum function
$p_{cl}(x,E)$, we state the physical condition which completes the
definition of the QMF $p(x,E)$ as:
\begin{equation}
p(x,E) \stackrel{\hbar \rightarrow 0}\longrightarrow p_{cl}(x,E)
\label{2.3a}
\end{equation}
Requirement (\ref{2.3a} ) has two interpretations: (1) as a form
of the correspondence principle and (2) as a boundary condition on
$p(x,E)$.

The boundary condition given above is easy to implement only for
very simple potentials because $p_{cl}(x,E)=\sqrt{E-V(x)}$ will in
general have several branch points, and the correct branch need to
be selected.  For this reason, we will impose other condition to
select the solution.  Several possibilities exists for an
alternate condition.  For example, the square integrability of the
bound state wave-function is one such requirement.  Some other
conditions, useful in the context of super symmetric potentials 
models, can also be written down [4].

%XXXXXXXXXXXXXXXXXXXXXXXXXXXXXXXXXXXXXXXXXXXXXXXXXXXXXXXXXXXXXXXXXXXXXXXXXX

\subsection{Exact Quantization Condition}

Having introduced the QMF $p(x,E)$, we define the quantum action
variable by generalizing the classical definition.  The classical
action variable can be defined as the integral

\[ \frac{1}{2\pi}\oint_{C} {p_{cl}(x,E)}dx  \]
where the integral is around a closed contour $C$.  The contour
$C$ encloses the cut of $p_{cl}(x,E)$  which runs between the
turning points $x_{1}$ and $x_{2}$.

Following the above definition, we define the quantum action
variable by
\begin{equation}
J=J(E)\equiv \frac{1}{2\pi}\oint_{C} {p(x,E)}dx \label{2.3b}
\end{equation}
where $p(x,E)$ is the quantum momentum function, and $C$ is the
contour defined immediately above.

The definition (\ref{2.3b}) connects the action-variable
eigen-value $ J $ to the energy eigen-value $ E $.  In order to
use (\ref{2.3b}) it is necessary to obtain the eigen-values $ J $.
Equation (\ref{2.3a}) and (\ref{2.3b}) imply that $p(x,E)$ has
poles of residue $-i\hbar$ on $Re$ $ x $-axis between the turning
points $x_{1}$ and $x_{2}$. For the ground state, first excited
state, second excited state
 \ $\cdots$, $p(x,E)$ has zero, one, two  \ $\cdots$, poles
respectively in the potential well.  The number of poles of
$p(x,E)$ in the potential well gives the excitation level of the
system. Since these poles of $p(x,E)$ are enclosed by the contour
$C$, we have
\begin{equation}
J=n\hbar = J(E)  \label{2.3c}
\end{equation}
where $ n= 0,1,2, \  \cdots $, and $E$ is the energy eigen-value
that is correlated with the values of $n\hbar$ for $J$.

Equation (\ref{2.3c}) can be inverted.  Thus one has
\begin{equation}
J=J(E) \  \rm{or}  \qquad  E = E(J)
\end{equation}

%XXXXXXXXXXXXXXXXXXXXXXXXXXXXXXXXXXXXXXXXXXXXXXXXXXXXXXXXXXXXXXXXXXXXXXXXXXX
%XXXXXXXXXXXXXXXXXXXXXXXXXXXXXXXXXXXXXXXXXXXXXXXXXXXXXXXXXXXXXXXXXXXXXXXXXXXXXXX
\section{Connection with
Schr$\ddot{o}$dinger Equation}

To bring out an equivalence between the QHJ equation and the
Schr$\ddot{o}$dinger equation, Leacock defines the wave-function
as
\begin{equation}
\psi (x,E) \equiv \exp\left(\frac{i}{\hbar}W(x,E)\right)
\end{equation}
The wave-function $\psi (x,E)$ satisfies the correct
Schr$\ddot{o}$dinger equation and the appropriate physical
boundary conditions. That the wave-function satisfies the correct
Schr$\ddot{o}$dinger equation can be seen from the above
definition and the QHJ equation for $W(x,E)$. The wave-function
for the bound states in one dimension has nodes whose number
increases with energy; the wave-function for the $n^{th}$ excited
state have $n$ nodes in the classical region. Since the QMF is
\begin{equation}
p(x,E) = -i\hbar \frac{\psi ^{\prime}(x)}{\psi (x)}
\end{equation}
these nodes are reflected as poles in the QMF and the residue of
QMF at each pole is $-i\hbar $.  Therefore, if we take a contour
integral  $$ \oint {p(x,E)}dx, $$ along a contour enclosing the
poles of QMF corresponding to the nodes of the wave function,
 we will have
\begin{equation}
\oint {p(x,E)}dx =2\pi i \ \mbox{\rm  [sum \  of  \  the  \
residues]} =nh
\end{equation}
This quantization condition along with the singularities and
knowledge of residues of QMF is sufficient for obtaining energy
eigen-values for exactly solvable models.  For other models,
approximation schemes can be developed.

%XXXXXXXXXXXXXXXXXXXXXXXXXXXXXXXXXXXXXXXXXXXXXXXXXXXXXXXXXXXXXXXXXXXXXX
%XXXXXXXXXXXXXXXXXXXXXXXXXXXXXXXXXXXXXXXXXXXXXXXXXXXXXXXXXXXXXXXXXXXXXXXXX

\section{Singularities of QMF}

The QHJ equation(\ref{2.2c}) is of Riccati form. If $V(x)$ has a
singular point, in the complex plane, $p(x,E)$ will also have
singular point at that location.  Such singular points are known
as fixed singular points, and will be present in every solutions.
On the other hand, other types of singular points with locations
depending on the initial conditions, may also be present.  These
singular points are known as moving singular points.  A well known
theorem states that, the moving singular points of solutions of
Riccati equation can only be poles. This pole will corresponds to
a zero of the wave-function.  Such a pole can only be a simple
pole with residue $-i\hbar$.  In fact if we substitute, assuming
$b\neq 0$,
\begin{equation}
p(x,E)\sim \frac{b}{(x-x_{0})^{r}}+\cdots
\end{equation}
in the QHJ equation
\begin{equation}
p^{2}(x,E)-i\hbar p^{\prime}(x,E)-[E-V(x)]=0
\end{equation}
and if the potential is not singular at $x=x_{0}$ then $r$ must be
equal to one and $b=-i\hbar$.  Thus the residues at each moving
pole must be $-i\hbar$.  This fact will be utilized throughout the
thesis.

In the next section we show how to calculate the eigen-values by
taking Morse oscillator as an example.

%XXXXXXXXXXXXXXXXXXXXXXXXXXXXXXXXXXXXXXXXXXXXXXXXXXXXXXXXXXXXXXXXXXXXXXXXXXXXXX
%XXXXXXXXXXXXXXXXXXXXXXXXXXXXXXXXXXXXXXXXXXXXXXXXXXXXXXXXXXXXXXXXXXXXXXXXXXXXX
\section{Energy Spectrum of Morse Oscillator}

The potential energy of the Morse oscillator is
\begin{equation}
 V(x)=A^{2}+B^{2}e^{-2\alpha x}-2B(A+\frac{\alpha}{2})e^{-\alpha
x}
\end{equation}
with the super potential
\begin{equation}
W(x)=A-Be^{-\alpha x}
\end{equation}
and $s=\frac{a}{\alpha}$

The quantum Hamilton-Jacobi equation is given by $(\hbar=1=2m)$
\begin{equation}
p^{2}(x,E)-ip^{\prime}(x,E) -\left[E-A^{2}-B^{2}e^{-2\alpha
x}+2B(A+\frac{\alpha}{2})e^{-\alpha x} \right]=0
\end{equation}
We effect a transformation to a new variable
\begin{equation}
y=\frac{2B}{\alpha}e^{-\alpha x}
\end{equation}
The quantum Hamilton-Jacobi equation in the new variable is
\begin{equation}
\tilde{p}^{2}(y,E)+i\alpha y \tilde{p}^{\prime}(y,E)- \left[E -
A^{2} -\frac{\alpha ^{2}}{4}y^{2} + (A + \frac{\alpha}{2})\alpha
y\right] = 0  \label{2.6a}
\end{equation}
where $\tilde{p}(y)\equiv p(x(y))$. We define $\phi(y,E)$ by
\begin{equation}
\tilde{p}(y,E)=i\alpha y\phi(y,E)
\end{equation}
Therefore (\ref{2.6a}) transforms to
\begin{equation}
 (\phi + \frac{1}{2y})^{2} + \phi^{\prime}- \frac{1}{4y^{2}}
+\frac{1}{\alpha^{2}y^{2}}[E  - A^{2} -\frac{y^{2}\alpha ^{2}}{4}
+ (A + \frac{\alpha}{2})y\alpha] = 0  \label{2.6b}
\end{equation}
Let

\begin{equation}
\chi(y,E) = \phi(y,E) + \frac{1}{2y}
\end{equation}
Therefore (\ref{2.6b}) transforms to
\begin{equation}
\chi^{2}  + \chi^{\prime} +
\frac{1}{4y^{2}}+\frac{1}{\alpha^{2}y^{2}}[E  - A^{2}
-\frac{y^{2}\alpha ^{2}}{4} + (A + \frac{\alpha}{2})y\alpha] = 0
\label{2.6c}
\end{equation}
$\chi $ has poles at $ y=0 $ and there are moving poles between
the classical turning points .  We assume that there are no more
poles in the complex plane other than a pole of finite order at
infinity.

{\it{\bf{Residue at the fixed pole $y=0$}}}: For $ y=0 $ we define
\begin{equation}
\chi = \frac{b_{1}}{y} + a_{0} + a_{1}y + \ldots  \label{2.6d}
\end{equation}
Using (\ref{2.6d}) in (\ref{2.6c}) and equating the coefficient of
$\frac{1}{y^{2}}$, yields
\begin{equation}
 b_{1} = \frac{1}{2} \ [1 \pm i \frac{2}{\alpha} \sqrt{|E -
A^{2}|}\ ]  \label{2.6e}
\end{equation}
The residue has two values, and the correct value is selected by
imposing the condition given below using the super potential viz.,

\[ \lim_{E \to 0} p(x,E) = i\sqrt{2m}  \ W(x) \]
In the y variable the above becomes (set $2m=1 $)
\[ \lim_{E \to 0} \tilde{p}(y,E) = i\tilde{W}(y) \]
which yields the value of $ b_{1}$ as in the $\lim_{E \to 0}$ as
\[ b_{1} = \frac{A}{\alpha}+\frac{1}{2}
\]

Hence the correct sign of $ b_{1}  $ is to choose the negative
sign in (\ref{2.6e}) and the hence the value of $ b_{1}  $ is

\begin{equation}
 b_{1} = \frac{1}{2} \ [1 - i \frac{2}{\alpha} \sqrt{E -
A^{2}|}\ ]
\end{equation}

{\it{\bf{Residue at $y=\infty$}}}: Now we determine the residue
for the pole at infinity, for which we effect a transformation
given by
\begin{equation}
y=\frac{1}{t}
\end{equation}
With $\tilde{\chi}(t)\equiv\chi(1/t)$ (\ref{2.6c}) transforms to
\begin{equation}
\tilde{\chi} ^{2}(t) -t^{2}\tilde{\chi}
^{\prime}(t)+\frac{1}{4}t^{2}+\frac{1}{\alpha^{2}}
\left[(E-A^{2})t^{2}-\frac{\alpha^{2}}{4}+(A+\frac{\alpha}{2})\alpha
t \right] =0  \label{2.6f}
\end{equation}
We assume an expansion for $\tilde{\chi} (t) $ as
\begin{equation}
\tilde{\chi} (t) = d_{0}+d_{1}t +d_{2}t^{2}+\cdots  \label{2.6g}
\end{equation}
The residue of $\tilde{\chi}(t)$ at $t=0$ is obtained from the
integral
\begin{equation}
\frac{1}{2\pi}\oint {p(x,E)}dx
\end{equation}
which in the $t$ variable yields the residue to be $d_{1}$.  To
determine the residue $d_{1}$ we use (\ref{2.6g}) in (\ref{2.6f}).
Therefore (\ref{2.6f}) transforms to
\begin{equation}
[d_{0}+d_{1}t +d_{2}t^{2}+\cdots]^{2}-t^{2}[d_{1} +2d_{2}t+\cdots]
+\frac{1}{4}t^{2}+\frac{1}{\alpha^{2}}
\left[(E-A^{2})t^{2}-\frac{\alpha^{2}}{4}+(A+\frac{\alpha}{2})\alpha
t \right] =0
\end{equation}
Equating the constant term on both sides gives
\begin{equation}
d_{0}=\pm \frac{1}{2}
\end{equation}
Equating the  power of $t$ on both sides gives
\begin{equation}
d_{1}=-\frac{1}{2\alpha d_{0}}\left[A+\frac{\alpha}{2}\right]
\label{2.6h}
\end{equation}
The correct sign for $d_{0}$ is chosen by the condition of square
integrability on the wave function viz.,
\[\psi (x) =\exp \left(i\int {p(x,E)}dx\right)
\]
The above integral is bounded at infinity only if
$d_{0}=-\frac{1}{2}$

Summarizing we have the result that for large $t$, $\tilde{\chi}$ is given
by (\ref{2.6g}) where $d_{0}=-\frac{1}{2}$, $d_{1}$ by
(\ref{2.6h})

\underline{\bf{The Quantization Rule and Eigen-values}}:

We shall now obtain the eigen-values by enforcing the quantization
rule

\[
J(E)=\frac{1}{2\pi}\oint_{C}{p(x,E)}dx = n\hbar
\]
where $C$ is a contour enclosing the part of real axis between the
turning points in the complex $x$-plane.  Changing the variable to
$y=\frac{2B}{\alpha}\exp\left(-\alpha x \right) $, the
corresponding quantization condition in the $y$-plane becomes

\[
J(E)=\frac{i}{2\pi}\oint_{C ^{\prime}}{(\tilde{\chi} - \frac{1}{2y})} dy =
n\hbar
\]
where $C^{\prime}$ is the image in the $y$-plane of the contour
$C$ in the $x$-plane, but with anti-clockwise orientation which
compensates for the negative sign coming from the derivative.  In
addition to the moving poles, $\tilde{p}(y,E)$ has a fixed pole at $y=0$.
Let $\gamma_{1}$ be a small circle enclosing the singular point
$y=0$, and $\Gamma_{R}$ is a circle of large radius $R$ such that
it encloses all the singularities of $p(x,E)$.  See fig(2.1).

Hence
\begin{equation}
I_{\Gamma_{R}} =J(E) + I_{\gamma_{1}} \label{2.6i}
\end{equation}
where $I_{\gamma_{1}}$ is the contour integral for the contour
$\gamma_{1}$ enclosing the pole  $y=0$ and $I_{\Gamma_{R}}$ is the
contour integral for the contour $\Gamma_{R}$.

We have evaluated the contour integral $I_{\gamma_{1}}$ and its
value is $-b_{1}$. The value of $J(E)$ is $-n$.

\begin{center}
\includegraphics*{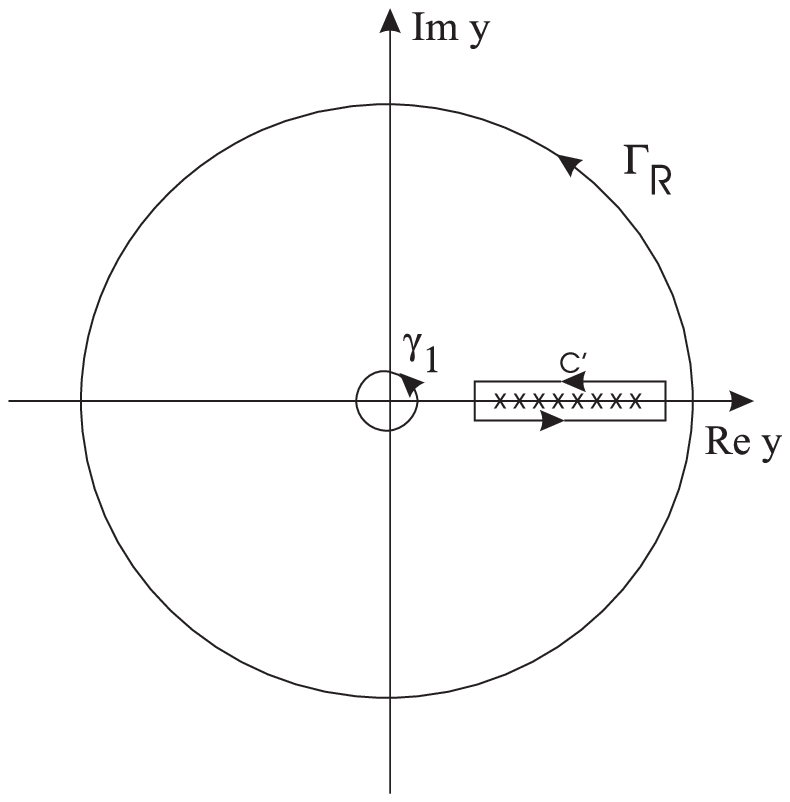}
%\caption{Fig. 2. 1}
\end{center}
%\newcommand{\hide}[1]{}
%
%\hide{ \setlength{\unitlength}{1mm}
%\begin{picture}(100,110)(0,0)
%\put(25,5){\epsfxsize=9cm\epsfbox{kgg1.eps}}
%\put(25,5){\raisebox{-5mm}{\makebox(90,0){Contour For Morse
%Oscillator}}}
%\end{picture}
%}
%
%
%
%\begin{figure}[bht]
%\begin{center}
%\epsfxsize=9cm \epsfbox{kgg1.eps} \caption{Contour for Evaluation
%of Action Integral }
%\end{center}
%\end{figure}
%
%
For evaluating the contour integral $I_{\Gamma_{R}}$ we make a
transformation of variable by $y=\frac{1}{t}$ and hence the
contour deforms to a new contour  $\Gamma_{r}$ which encloses the
singular point at $y=\infty$ or $t=0$.  The value of this contour
integral $I_{\Gamma_{r}}$ has been evaluated and is $-d_{1}$.

Hence (\ref{2.6i}) transforms to
\begin{equation}
I_{\Gamma_{r}} =J(E) + I_{\gamma_{1}}
\end{equation}

Hence to obtain the energy spectra of the Morse oscillator, we
equate the residue of the fixed poles and the moving poles to
those at infinity.    Hence we have
\begin{equation}
-b_{1}-n=-d_{1}
\end{equation}
Substituting the values of $b_{1}$ and $d_{1}$ we get
\begin{equation}
\frac{1}{2}  \big(1 - i \frac{2}{\alpha} \sqrt{|E - A^{2}|}\big)
+n=-\frac{1}{2\alpha d_{0}}\left[A+\frac{\alpha}{2}\right]
\end{equation}
which on simplification gives the desired result for energy
spectrum as
\begin{equation}
E=A^{2} - (A-n\alpha)^{2}
\end{equation}

%xxxxxxxxxxxxxxxxxxxxxxxxxxxxxxxxxxxxxxxxxxxxxxxxxxxxxxxxxxxxxxxxxxxxxxxx
%XXXXXXXXXXXXXXXXXXXXXXXXXXXXXXXXXXXXXXXXXXXXXXXXXXXXXXXXXXXXXXXXXXXXXXXX
%xxxxxxxxxxxxxxxxxxxxxxxxxxxxxxxxxxxxxxxxxxxxxxxxxxxxxxxxxxxxxxxxxxxxxxxx
\chapter{CACULATION OF WAVE-FUNCTION FOR ES MODELS}

\newcommand{\csch}{\mbox{\rm csch}}
\newcommand{\sech}{\mbox{\rm sech}}

In this chapter we apply the QHJ formalism outlined in the
previous chapter, to find bound state wave-functions for several
exactly solvable potential problems in one dimension.  We will
show that, by making use of elementary theorems in complex
variables, the form of QMF can be determined completely and hence
the bound state wave-functions are easily obtained.  To determine
the form of QMF we begin with the QHJ equation $(\hbar=1=2m)$
\[
 p^{2}(x,E)-ip^{\prime}(x,E)-[E-V(x)]=0
\]
where p(x,E) is the QMF continued in the complex x-plane, and is
related to the wave-function by
\[
 p(x,E)=-i\frac{\psi ^{\prime}(x)}{\psi (x)}
\]
The zeros of the wave-function will appear as poles in the QMF.
According to the well known theorems about the nodes of
wave-function, the $n^{th}$ excited state corresponds to $n$ zeros
on the real line, and there will be corresponding $n$ ($moving$)
poles in the QMF and the residue at each pole will be $ -i$ as has
been discussed in chapter 2.  In addition to these moving poles,
there are fixed poles corresponding to the singularities of the
potential.  {\it {\bf{We will make an assumption that QMF has no
other singularities in the finite complex plane}}}.  The QMF turns
out to be meromorphic, and to fix its form one needs to know the
behavior of QMF for large $x$ in the complex $x$-plane.  This
information can be easily read from the QHJ equation and hence the
form of QMF can then be fixed completely.  In the next section we
show how this strategy works for the harmonic oscillator.  In the
remaining sections of this chapter we give the details of the
calculation of the bound state wave-functions for harmonic
oscillator, Morse oscillator, Poschl Teller, Eckart potentials
and hydrogen atom. For these potentials a change of variable
becomes necessary and we always try to bring the QHJ equation in
the new variable to a  form  as  the above equation.  We also
mention that several other potentials have been studied [5] and
the bound state wave-functions in each case agree with the known
results [8].
%XXXXXXXXXXXXXXXXXXXXXXXXXXXXXXXXXXXXXXXXXXXXXXXXXXXXXXXXXXXXXXXXXXXXXX
%xxxxxxxxxxxxxxxxxxxxxxxxxxxxxxxxxxxxxxxxxxxxxxxxxxxxxxxxxxxxxxxxxxxxxx
\section{Harmonic Oscillator}

The potential energy of the harmonic oscillator is
\begin{equation}
 V(x)=\frac{1}{2}m\omega^{2}x^{2}
\end{equation}
The quantum Hamilton-Jacobi equation is given by ($\hbar=1=2m$)
\begin{equation}
p^{2}(x,E)-ip^{\prime}(x,E)-[E-\frac{1}{4}\omega^{2}x^{2}]=0
\label{3.1a}
\end{equation}

The QMF $p(x,E)$ has $n$ poles corresponding to the zeros of the
wave-function, and residue at each of these poles is $-i$.  It can
be proved that $p(x,E)$ has no other poles except at infinity [1].

For large $x$
\begin{equation}
p(x,E)\approx \pm \frac{1}{2}i\omega x
\end{equation}
and we write
\begin{equation}
p(x,E)\approx \pm \frac{1}{2}i\omega x +Q(x)
\end{equation}
where $Q(x)$ is  to be determined.

The sign of  $p(x,E)$ is determined by the condition of square
integrability of the wave-function.

The wave-function is expressed as
\begin{equation}
\psi(x)=\exp\left(i\int {p(x,E)}dx \right)
\end{equation}
When the above value of $p(x,E) $ is substituted in the equation
of wave-function, the wave-function is bounded at at large $x$ if
we choose the positive sign of $\frac{1}{2}i \omega x $.

Hence we write the quantum momentum function $p(x,E)$ as
\begin{equation}
p(x,E)= \sum_{k=1}^{n} \frac{-i}{x-x_{k}}+ \frac{1}{2}i \omega x +
\phi(x)
\end{equation}
where $x_{1},x_{2}, \cdots, x_{n}$ are the location of $n$ poles
on the $x$-axis and $\phi(x)$ is analytic  every where and bounded
at infinity. Therefore  Liouville's theorem tells us, it has to be
a constant. Hence let $\phi(x)=c$ a constant.  Hence the above
equation becomes
\begin{equation}
p(x,E)= \sum_{k=1}^{n} \frac{-i}{x-x_{k}}+ \frac{1}{2}i \omega x +
c \label{3.1b}
\end{equation}
Substituting (\ref{3.1b}) in (\ref{3.1a}) we have
\begin{equation}
\left[\sum_{k=1}^{n} \frac{-i}{x-x_{k}}+ \frac{1}{2}i \omega x +
c\right]^{2} + \sum_{k=1}^{n} \frac{1}{(x-x_{k})^{2}}
-[E-\frac{1}{4}\omega^{2}x^{2}]=0  \label{3.1c}
\end{equation}
Equating the power of $x$, we have $c=0$. Equating the constant
term to zero on both sides in equation (\ref{3.1c}) gives
\begin{equation}
E=(n+\frac{1}{2})\omega
\end{equation}
which is the well known expression for energy of the harmonic
oscillator in our notation ($2m=1$)

The sum of moving pole terms $$\sum_{k=1}^{n} \frac{-i}{x-x_{k}}$$
can be expressed as $\frac{P^{\prime}(x)}{P(x)}$  where  $P(x)$ is
the polynomial
\begin{equation}
P(x)=\prod_{k=1}^{n} (x-x_{k}).  \label{3.1d}
\end{equation}
The QMF (\ref{3.1b}) can be expressed as
\begin{equation}
p(x,E)=-i\frac{P^{\prime}(x)}{P(x)}+\frac{1}{2}i \omega x,
\label{3.1e}
\end{equation}
Using (\ref{3.1e}) in (\ref{3.1a}) and on simplification yields
\begin{equation}
P^{{\prime}{\prime}}(x) -\omega x P^{\prime}(x)+n\omega P(x)=0
\label{3.1f}
\end{equation}
We effect a transformation $\xi=\alpha x $ where $
\alpha^{2}=\frac{\omega}{2}$.  Hence equation (\ref{3.1f}) changes
to
\begin{equation}
P^{{\prime}{\prime}}(\xi) - 2\xi P^{\prime}(\xi)+2n P(\xi)=0
\end{equation}
The above equation resembles the well known Hermite differential
equation.  Hence on comparison,  we get
\begin{equation}
P(\xi)\equiv H_{n}(\alpha x)
\end{equation}
and is the Hermite polynomial.

The wave-function is expressed as
\[
\psi(x)=\exp\left[i\int {p(x,E)}dx \right]=\exp\left[i\int
{(-i\frac{P^{\prime}(x)}{P(x)}+i\frac{1}{\sqrt{2}}\omega x) }dx
\right].
\]
and hence we have
\begin{equation}
\psi(x)=H_{n}(\alpha x) \exp\left( -\frac{1}{4}\omega
x^{2}\right).
\end{equation}
This is the desired wave-function for the harmonic oscillator.

%XXXXXXXXXXXXXXXXXXXXXXXXXXXXXXXXXXXXXXXXXXXXXXXXXXXXXXXXXXXXXXXXXXXXXXXX
%XXXXXXXXXXXXXXXXXXXXXXXXXXXXXXXXXXXXXXXXXXXXXXXXXXXXXXXXXXXXXXXXXXXXXXX
\section{Morse Oscillator}

The potential energy of the Morse oscillator is
\begin{equation}
 V(x)=A^{2}+B^{2}e^{-2\alpha x}-2B(A+\frac{\alpha}{2})e^{-\alpha
x}
\end{equation}
with the super potential
\begin{equation}
W(x)=A-Be^{-\alpha x}
\end{equation}
and
\begin{equation}
s=\frac{A}{\alpha}.
\end{equation}

The quantum Hamilton-Jacobi equation is given by $(\hbar=1=2m)$
\begin{equation}
p^{2}(x,E)-ip^{\prime}(x,E) -\left[E-A^{2}-B^{2}e^{-2\alpha
x}+2B(A+\frac{\alpha}{2})e^{-\alpha x} \right]=0
\end{equation}
We effect a transformation to the variable
\begin{equation}
y=\frac{2B}{\alpha}e^{-\alpha x}
\end{equation}
The quantum Hamilton-Jacobi equation in the new variable is
\begin{equation}
\tilde{p}^{2}(y,E)+i\alpha y \tilde{p}^{\prime}(y,E)- \left[E -
A^{2} -\frac{\alpha ^{2}}{4}y^{2} + (A + \frac{\alpha}{2})\alpha
y\right] = 0  \label{3.2a}
\end{equation}
We define
\begin{equation}
\tilde{p}(y,E)=i\alpha y\phi(y,E)
\end{equation}
Then (\ref{3.2a}) transforms to
\begin{equation}
 (\phi + \frac{1}{2y})^{2} + \phi^{\prime}- \frac{1}{4y^{2}}
+\frac{1}{\alpha^{2}y^{2}}[E  - A^{2} -\frac{y^{2}\alpha ^{2}}{4}
+ (A + \frac{\alpha}{2})y\alpha] = 0  \label{3.2b}
\end{equation}
Let

\begin{equation}
\chi(y,E) = \phi(y,E) + \frac{1}{2y}
\end{equation}
Therefore (\ref{3.2b}) transforms to
\begin{equation}
\chi^{2} + \frac{1}{4y^{2}} + \chi^{\prime}
+\frac{1}{\alpha^{2}y^{2}}[E  - A^{2} -\frac{y^{2}\alpha ^{2}}{4}
+ (A + \frac{\alpha}{2})y\alpha] = 0   \label{3.2c}
\end{equation}

This equation suggests that $\chi$ has a pole at $y=0$.  It will
also have $n$ moving poles corresponding to the nodes of the
wave-function.  We assume that there are no other poles in the
finite complex plane.  For large $y$ the behavior of $\chi$ has
already been obtained in section (2.6) and is seen to be bounded
for large $y$.  Hence we get using Liouville's theorem
\begin{equation}
\chi =
\frac{b_{1}}{y}+\sum_{k=1}^{n}\left(\frac{1}{y-y_{k}}\right) +c
\label{3.2d}
\end{equation}
where $b_{1}$ and $c$ are constants to be fixed.  The residue of
$\chi$ at $y=0$ is $b_{1}$ and has been obtained in section
(2.6) and
\begin{equation}
b_{1}=\frac{A}{\alpha}+\frac{1}{2}
\end{equation}
We write, once again,
\[\sum_{k=1}^{n}\frac{1}{y-y_{k}}=\frac{P^{\prime}(y)}{P(y)}
\]
where
\[ P(y) =\prod_{k=1}^{n}(y-y_{k})
\]
Substituting (\ref{3.2d})  in  (\ref{3.2c}) gives
\begin{equation}
\frac{P^{{\prime}{\prime}}}{P} + 2
\frac{b_{1}}{y}\frac{P^{\prime}}{P} + 2\frac{P^{\prime}}{P}c +
2\frac{b_{1}}{y}c + c^{2} - \frac{1}{4} + \frac{1}{\alpha y}(A +
\frac{\alpha}{2}) = 0  \label{3.2e}
\end{equation}
In order to proceed further we look at the behavior of each term
for large $y$.  Using the leading terms
\[
\frac{P^{{\prime}{\prime}}(y)}{P(y)}\sim \frac{n(n-1)}{y^{2}} \ ,
\qquad  \frac{P^{\prime}(y)}{P(y)}\sim \frac{n}{y}
\]
in equation (\ref{3.2e}) and equating the constant term on both
sides gives, $ c = \pm \frac{1}{2}. $  The correct sign for $c $
is chosen by the condition of square integrability on the wave
function which fixes $c=-\frac{1}{2}. $

Comparing the coefficient of $ \frac{1}{y} $ for large y on both
sides  we get
\begin{equation}
2b_{1}c+2nc+\left(A+\frac{\alpha}{2}\right)\frac{1}{\alpha}=0
\end{equation}
which on using the values of $b_{1}$ and $c$ and on simplification
gives the energy eigen-value
\begin{equation}
E = A^{2} -(A -n\alpha)^{2}
\end{equation}
Substituting the value of $ b_{1} $and $c $ in equation
(\ref{3.2e}) we have
\begin{equation}
 yP^{{\prime}{\prime}}(y) + \{1-y +2(s-n) \}P^{\prime}(y) +nP(y) = 0
\end{equation}
Compare this with the standard Laguerre differential equation
\[
xy^{{\prime}{\prime}} + (\beta +1 -x)y^{\prime} +ny = 0
\]
we have  $P(y) \equiv L_{\beta}^{n}(y)$.

The wave-function for the Morse oscillator is given by
\begin{equation}
\psi(x)=\exp\left(i\int {p(x,E)}dx\right)
\end{equation}
In terms of $y$ variable we have
\begin{equation}
\psi(y)=\exp\left(i\int
{\left[\frac{b_{1}}{y}+\frac{P^{\prime}(x)}{P(x)}-\frac{1}{2}-\frac{1}{2y}\right]}dy
\right)
\end{equation}
On integrating and simplifying we get
\begin{equation}
\psi_{n}(y)= y^{s-n}\exp(-\frac{1}{2}y)P(x)
\end{equation}
Replacing the  value of $P(y)$ we have
\begin{equation}
\psi_{n}(y)= y^{s-n}\exp(-\frac{1}{2}y)L_{\beta}^{n}(y)
\end{equation}

%XXXXXXXXXXXXXXXXXXXXXXXXXXXXXXXXXXXXXXXXXXXXXXXXXXXXXXXXXXXXXXXXXXXXXX
%XXXXXXXXXXXXXXXXXXXXXXXXXXXXXXXXXXXXXXXXXXXXXXXXXXXXXXXXXXXXXXXXXXXXXX
\section{Poschl-Teller Potential}

The  Poschl-Teller  potential is
\begin{equation}
V(x)=A^{2}+(B^{2}+A^{2}+A\alpha)\ \csch^{2}\alpha x
-B\,(2A+\alpha)\coth \alpha x \ \csch\alpha x
\end{equation}
with the super potential given by
\begin{equation}
W(x)=A\coth \alpha x - B\ \csch\, \alpha x  \  (A < B)
\end{equation}
and
\begin{equation}
s= \frac{A}{\alpha} \ ,\qquad  \lambda = \frac{\beta}{\alpha}
\end{equation}

The quantum Hamilton-Jacobi equation is given by $(\hbar=1=2m)$
\[
p^{2}(x,E)-ip^{\prime}(x,E)
\]
\begin{equation}
-\left[E- A^{2}-(B^{2}+A^{2}+A\alpha)\ \csch^{2}\alpha x
+B(2A+\alpha)\coth \alpha x\ \csch\,\alpha x\right]=0
\end{equation}
We effect a transformation to a new variable
\begin{equation}
y = \cosh \alpha x
\end{equation}
The quantum Hamilton-Jacobi equation in the new variable becomes
\[
\tilde{p}^{2}(y,E)-i\alpha \sqrt{y^{2}-1}\tilde{p}^{\prime}(y,E)
\]
\begin{equation}
-\left[E- A^{2}-(B^{2}+A^{2}+A\alpha)\frac{1}{y^{2}-1}
+B(2A+\alpha)\frac{y}{y^{2}-1} \right]=0 \label{3.3a}
\end{equation}
We define

\begin{equation}
\tilde{p}(y,E)=-i\alpha \sqrt{y^{2}-1}\phi(y).
\end{equation}
Therefore (\ref{3.3a}) transforms to
\[
(\phi +\frac{1}{2}\frac{y}{y^{2}-1})^{2}+\phi
^{\prime}-\frac{1}{4}\frac{y^{2}}{(y^{2}-1)^{2}}
\]
\begin{equation}
+\frac{1}{\alpha ^{2}(y^{2}-1)}\left[E-
A^{2}-(B^{2}+A^{2}+A\alpha)\frac{1}{y^{2}-1}
+B(2A+\alpha)\frac{y}{y^{2}-1} \right]=0  \label{3.3b}
\end{equation}
Let
\begin{equation}
\chi =\phi + \frac{1}{2}\frac{y}{y^{2}-1}.
\end{equation}
Therefore (\ref{3.3b}) transforms to

\[
\chi ^{2} + \chi^{\prime}+\frac{3}{4}\frac{y^{2}}{(y^{2}-1)^{2}}
-\frac{1}{2}\frac{1}{y^{2}-1}
\]
\begin{equation}
+\frac{1}{\alpha ^{2}(y^{2}-1)}\left[E-
A^{2}-(B^{2}+A^{2}+A\alpha)\frac{1}{y^{2}-1}
+B(2A+\alpha)\frac{y}{y^{2}-1} \right]=0 \label{3.3c}
\end{equation}
$ \chi $ has poles at  $ y= \pm 1 $ and there are moving poles
between the classical turning points.  We assume that there are no
more poles in the complex line.  We now determine the residue at
each of these poles.

For $ y=+1 $ , we define
\begin{equation}
\chi =\frac{b_{1}}{y-1}+a_{0}+a_{1}(y-1)+\cdots  \label{3.3d}
\end{equation}
Using (\ref{3.3d}) in (\ref{3.3c}) and equating the coefficient of
$\frac{1}{(y-1)^{2}} $ yields
\begin{equation}
b_{1}=\frac{1}{2}\left[1 \pm
\frac{1}{2\alpha}[2(B-A)-\alpha]\right]
\end{equation}
The correct value of  $ b_{1}  $ is selected by imposing the
condition
\[
\lim_{E \to 0} \mbox{\rm Res } \tilde{p}(y,E)=i  \mbox{\rm Res} \
\tilde{W}(y,E)
\]
on super potential as explicitly shown for Morse oscillator gives
\begin{equation}
b_{1}=\frac{1}{4}-\frac{1}{2\alpha}(A-B)
\end{equation}

For $ y=-1 $ , we define
\begin{equation}
\chi
=\frac{b_{1}^{\prime}}{y+1}+a_{0}^{\prime}+a_{1}^{\prime}(y+1)+\cdots
\label{3.3e}
\end{equation}
Using (\ref{3.3e}) in (\ref{3.3c}) and equating the coefficient of
$\frac{1}{(y+1)^{2}} $ and following the above procedure yields

\begin{equation}
b_{1}^{\prime}=\frac{1}{4}-\frac{1}{2\alpha}(A+B)
\end{equation}
The residue at a moving pole is seen from (\ref{3.3c}) to be 1.
Hence we arrive at the form

\begin{equation}
\chi = \frac{b_{1}}{y-1}+\frac{b_{1}^{\prime}}{y+1}
+\frac{P^{\prime}(y)}{P(y)}+c    \label{3.3f}
\end{equation}
for $\chi $ where $c$ is a constant to be determined

Substituting (\ref{3.3f}) in (\ref{3.3c}) gives the following
equation.

\[
c^{2}+\frac{2b_{1}b_{1}^{\prime}}{y^{2}-1}+\frac{2b_{1}^{\prime}}{y+1}\frac{P^{\prime}}{P}
+2\frac{P^{\prime}}{P}c+\frac{2b_{1}c}{y-1}+\frac{2b_{1}}{y-1}\frac{P^{\prime}}{P}
+\frac{2b_{1}^{\prime}}{y+1}c-\frac{1}{2}\frac{1}{y^{2}-1}+\frac{P^{{\prime}{\prime}}}{P}
\]
\begin{equation}
\frac{3}{8}\frac{1}{y^{2}-1}+\frac{1}{\alpha
^{2}}(E-A^{2})\frac{1}{y^{2}-1}+\frac{1}{2\alpha ^{2}
}(B^{2}+A^{2}+A\alpha)\frac{1}{y^{2}-1}=0  \label{3.3g}
\end{equation}
Now we look at different terms in equation (\ref{3.3g}) for large
$ y $ and equate their coefficient to zero.  Equating the constant
term to zero gives $ c= 0 $ . With $ c=0 $ the above equation
becomes
\[
\frac{P^{{\prime}{\prime}}}{P}+\frac{P^{\prime}}{P}\left[\frac{2b_{1}}{y-1}
+\frac{2b_{1}^{\prime}}{y+1}\right]+\frac{2b_{1}b_{1}^{\prime}}{y^{2}-1}-\frac{1}{2}\frac{1}{y^{2}-1}
\]
\begin{equation}
\frac{3}{8}\frac{1}{y^{2}-1}+\frac{1}{\alpha
^{2}}(E-A^{2})\frac{1}{y^{2}-1}+\frac{1}{2\alpha ^{2}
}(B^{2}+A^{2}+A\alpha)\frac{1}{y^{2}-1}=0   \label{3.3h}
\end{equation}

For large $ y , \  P(y)  $ behaves as $ P(y)\sim y^{n}+\cdots $
and $\frac{P^{{\prime}{\prime}}(y)}{P(y)}\sim \frac{n(n-1)}{y^{2}}
\ ,  \frac{P^{\prime}(y)}{P(y)}\sim \frac{n}{y} $

Using these in (\ref{3.3h}) and equating the coefficient of $
\frac{1}{y^{2}}$   gives
\[
2b_{1}b_{1}^{\prime}+2nb_{1}^{\prime}+2nb_{1}+n(n-1)+\frac{3}{8}+\frac{1}{\alpha
^{2}}(E-A^{2})+\frac{1}{2\alpha
^{2}}(B^{2}+A^{2}+A\alpha)-\frac{1}{2}=0
\]

and substituting the values of $ b_{1} $ and $b_{1}^{\prime} $
gives the expression for energy as

\begin{equation}
E=A^{2}-(A-n\alpha)^{2}
\end{equation}
Substituting the value of $ E , b_{1} $ and  $ b_{1}^{\prime}$ in
(\ref{3.3h}) we get
\begin{equation}
(1-y^{2})P^{{\prime}{\prime}}(y)+[(2s-1)y-2\lambda]P^{\prime}(y)+[n(n-2s)]P(y)=0
\end{equation}
The above equation resembles the standard Jacobi polynomial, such
that
\begin{equation}
P(y)\equiv P_{n}^{(\alpha , \beta)}(y)=P_{n}^{(\lambda - s
-\frac{1}{2},-\lambda - s - \frac{1}{2})}(y)
\end{equation}
The wave function for the Poschl Teller potential is obtained on
the same line as that for the Morse oscillator and is given by

\begin{equation}
\psi (y)= (y-1)^{\frac{(\lambda -s)}{2}}y+1)^{\frac{-(\lambda
+s)}{2}}P_{n}^{(\lambda - s -\frac{1}{2},-\lambda - s -
\frac{1}{2})}(y)
\end{equation}
which agrees well with the values given in literature. [5]

%XXXXXXXXXXXXXXXXXXXXXXXXXXXXXXXXXXXXXXXXXXXXXXXXXXXXXXXXXXXXXXXXXXXXXX
%XXXXXXXXXXXXXXXXXXXXXXXXXXXXXXXXXXXXXXXXXXXXXXXXXXXXXXXXXXXXXXXXXXXXXXXX
\section{Eckart Potential}

 The potential energy of the Eckart potential is
\begin{equation}
V(x)=A^{2}+\frac{B^{2}}{A^{2}}-2B\coth \alpha x+A(A-\alpha) \
\csch^{2}\alpha x
\end{equation}
with the super potential given by
\begin{equation}
W(x)=-A\coth \alpha x + \frac{B}{A}  \ ,\qquad (B>A^{2})
\end{equation}
and
\begin{equation}
s=\frac{A}{\alpha} \ ,\qquad \lambda = \frac{B}{\alpha ^{2}}  \ ,
\qquad  \alpha =\frac{\lambda}{n+s}
\end{equation}
The quantum Hamilton-Jacobi equation is given by ($\hbar=1=2m$)
\begin{equation}
p^{2}(x,E)-ip^{\prime}(x,E)-\left[E-A^{2}-\frac{B^{2}}{A^{2}}+2B\
\coth \alpha x-A(A-\alpha)\ \csch^{2} \alpha x\right]=0
\end{equation}
We effect a transformation by the variable
\begin{equation}
y=\coth \alpha x
\end{equation}
The quantum Hamilton-Jacobi equation in the new variable is
\begin{equation}
\tilde{p}^{2}(y,E)-i\alpha
(1-y^{2})\tilde{p}^{\prime}(y,E)-\left[E-A^{2}-\frac{B^{2}}{A^{2}}
+2By-A(A-\alpha)(y^{2}-1)\right]=0 \label{3.4a}
\end{equation}

We define
\begin{equation}
\tilde{p}(y,E)=-i\alpha (1-y^{2})\phi (y)
\end{equation}
Hence equation (\ref{3.4a}) simplifies to
\[
[\phi - \frac{y}{1-y^{2}}]^{2}+\phi ^{\prime} -
\frac{y^{2}}{(1-y^{2})^{2}}
\]
\begin{equation}
\frac{1}{\alpha ^{2}(1-y^{2})^{2}}
\left[E-A^{2}-\frac{B^{2}}{A^{2}}
+2By-A(A-\alpha)(y^{2}-1)\right]=0
\end{equation}
Let
\begin{equation}
\chi = \phi - \frac{y}{1-y^{2}}
\end{equation}
Therefore the above equation changes to
\[
\chi ^{2} + \chi ^{\prime} +
\frac{y^{2}}{(1-y^{2})^{2}}+\frac{1}{1-y^{2}}
\]
\begin{equation}
+\frac{1}{\alpha
^{2}(1-y^{2})^{2}}\left[E-A^{2}-\frac{B^{2}}{A^{2}}
+2By-A(A-\alpha)(y^{2}-1)\right]=0 \label{3.4b}
\end{equation}
$ \chi $ has poles at $ y=\pm 1$ and there are moving poles
between the classical turning points.  We assume that there are no
more poles in the complex plane.  We determine the residue at each
of these poles.

For $ y = +1 $, we define
\begin{equation}
\chi = \frac{b_{1}}{y-1}+a_{0}+a_{1}(y-1)+\cdots  \label{3.4c}
\end{equation}
Using (\ref{3.4c}) in (\ref{3.4b}) and equating the coefficient of
$\frac{1}{(y-1)^{2}} $  yields
\begin{equation}
b_{1}=\frac{1}{2}\left[1 \pm
\frac{1}{\alpha}\sqrt{\left(A-\frac{B}{A}\right)^{2}-E}\ \right]
\end{equation}
As the residue has two values, the correct value is selected by
imposing the condition on the super potential as done in Morse
oscillator and the correct value is
\begin{equation}
b_{1}=\frac{1}{2}\left[1 +
\frac{1}{\alpha}\sqrt{\left(A-\frac{B}{A}\right)^{2}-E}\ \right]
\end{equation}
Similarly the residue for  $y=-1  $ is determined and is given as
\begin{equation}
b_{1}^{\prime}=\frac{1}{2}\left[1 -
\frac{1}{\alpha}\sqrt{\left(A+\frac{B}{A}\right)^{2}-E}\ \right]
\end{equation}
We assume $ \chi $ to have the form
\begin{equation}
\chi=\frac{b_{1}}{y-1}+\frac{b_{1}^{\prime}}{y+1}+\frac{P^{\prime}(y)}{P(y)}+c
\label{3.4d}
\end{equation}
where $c$ is a constant to be determined

Using (\ref{3.4d}) in (\ref{3.4b}) and following the similar lines
as that of Morse and Poschl Teller potential one gets the value of
$ c=0 $ \ and the resulting equation becomes
\[
\frac{P^{{\prime}{\prime}}(y)}{P(y)}+\frac{{P^\prime}(y)}{P(y)}\left[\frac{2b_{1}}{y-1} +
\frac{b_{1}^{\prime}}{y+1}\right]+\frac{2b_{1}b_{1}^{\prime}}{y^{2}-1}
\]
\begin{equation}
-\frac{1}{2}\frac{1}{y^{2}-1}
-\frac{1}{2\alpha^{2}}\frac{1}{y^{2}-1}[E-A^{2}-\frac{B^{2}}{A^{2}}]-\frac{1}{\alpha
^{2}}\frac{1}{y^{2}-1}A(A-\alpha)=0  \label{3.4e}
\end{equation}
For large $y$ assuming $P(y)\sim y^{n}+\cdots$ and equating the
coefficient of $ \frac{1}{y^{2}} $ gives
\[
2b_{1}b_{1}^{\prime}+2b_{1}^{\prime}n+2b_{1}n+n(n-1)-\frac{1}{2}-\frac{1}{2\alpha^{2}}(E-A^{2}-\frac{B^{2}}{A^{2}})
-\frac{1}{\alpha^{2}}A(A-\alpha)=0
\]
Substituting the values of $b_{1}$ and $b_{1}^{\prime}$ gives the
energy expression as
\begin{equation}
E=A^{2}-(A+n\alpha)^{2}-\frac{B^{2}}{(A+n\alpha)^{2}}+\frac{B^{2}}{A^{2}}
\end{equation}
Using the values of $b_{1}, b_{1}^{\prime}$  and  $E$  in
(\ref{3.4e}) one gets the differential equation for Eckart
potential as
\begin{equation}
(1-y^{2})P^{{\prime}{\prime}}(y)+\left[-2\alpha-2(-n-s+1)y\right]P^{\prime}(y)+
(-2ns)P(y)=0 \label{3.4f}
\end{equation}
Equation (\ref{3.4f}) resembles the standard Jacobi polynomial and
\begin{equation}
P(y)\equiv P_{n}^{(\alpha , \beta)}(y)=P_{n}^{(s_{3},s_{4})}(y)
\end{equation}
The wave function for the Eckart Potential is obtained from
\[\psi (x)= e^{i\int {p(x,E)}dx} \]
and is given by
\begin{equation}
\psi (y) =
(y-1)^{\frac{s_{3}}{2}}(y+1)^{\frac{s_{4}}{2}}P_{n}^{(s_{3},s_{4})}(y)
\end{equation}
The values for energy and wave function agree with those found in
the literature.[5]

%XXXXXXXXXXXXXXXXXXXXXXXXXXXXXXXXXXXXXXXXXXXXXXXXXXXXXXXXXXXXXXXXXXXXXXXXX
%XXXXXXXXXXXXXXXXXXXXXXXXXXXXXXXXXXXXXXXXXXXXXXXXXXXXXXXXXXXXXXXXXXXXXXXXXX

\section{Hydrogen Atom}
In this section we obtain the bound state wave functions of the
radial part of the Schr$\ddot{o}$dinger equation $(\hbar = 2m =
1)$
\begin{equation}
\frac{d^{2}R}{dr^2} +
\frac{2}{r}\frac{dR}{dr}+\left(E+\frac{Ze^{2}}{r}-\frac{\lambda}{r^2}\right)R
= 0  \label{e21}
\end{equation}
where $\lambda = l(l+1)$. Using the transformation
$R(r)=\phi(r)/r$, the Schr$\ddot{o}$dinger equation becomes
\begin{equation}
\frac{d^{2}\phi}{dr^2}+(E+\frac{Ze^2}{r}-\frac{\lambda}{r^2})\phi
= 0.    \label{76}
\end{equation}
The QHJ equation in terms of
\begin{equation}
q = \frac{d}{dr}\ln(\phi(r))     \label{e75}
\end{equation}
is given by
\begin{equation}
q^2+\frac{dq}{dr}+\left(E+\frac{Ze^{2}}{r}-\frac{\lambda}{r^2}\right)=0.
\label{e23}
\end{equation}
The range of $r$ is from 0 to $\infty$, and the wave function
$\phi(r)$ should vanish at $r=0$. Thus $q$ has a fixed pole at
$r=0$, along with the $n$ moving poles with residue equal to one
on the real line. Like harmonic oscillator there are no other
singularities in the finite complex plane. Thus we can write $q$,
in a similar fashion as for harmonic oscillator, as
\begin{equation}
q(r) = \frac{P^{\prime}}{P} +\frac{b_{1}}{r}+C    \label{e24}
\end{equation}
where $b_{1}$ is the residue at $r=0$ which can be obtained by
doing a Laurent expansion of $q$ around the pole at the origin.
The two values of residues obtained are
\begin{equation}
b_{1}=-l ,\,\, b_{1}= l+1.     \label{e25}
\end{equation}
One chooses the right residue by using the square integrability
property of the wave function $\phi$ and obtain
\begin{equation}
b_{1}=l+1     \label{e26}
\end{equation}
as the right choice. Substituting (\ref{e24}) for $q$ in
(\ref{e23}) and expanding different terms of the resulting
equation for large $r$ and comparing the leading terms we get
\begin{equation}
C^2 = - E ,\,\,E= -\frac{Z^{2}e^{4}}{(2n^{\prime})^2} \label{e27}
\end{equation}
where $n^{\prime}= n+l+1$ and one is left with the differential
equation
\begin{equation}
rP^{\prime\prime}+2P^{\prime}\left(l+1-\frac{Ze^{2}r}{2n^{\prime}}\right)+\frac{(n^{\prime}-l-1)Ze^{2}}{n^{\prime}}P
  =0.  \label{e28}
\end{equation}
Now defining
\begin{equation}
\frac{Ze^{2}}{n^{\prime}}r=\rho    \label{e29}
\end{equation}
(\ref{e27}) becomes
\begin{equation}
\rho P^{\prime \prime}+((2l+1)+1-\rho)P^{\prime}+(n^{\prime}-l-1)P
= 0    \label{e30}
\end{equation}
which is the associated Laguerre differential equation where $P$
is the Laguerre polynomial denoted by $L$. The bound state wave
function obtained from (\ref{e24}) and (\ref{e75}) is
\begin{equation}
\psi_{n}(\rho)=\rho ^{l+1}\exp(-\rho
/2)L_{n^{\prime}+l}^{2l+1}(\rho)    \label{e31}
\end{equation}
which is seen to be identical with known correct answer.

%XXXXXXXXXXXXXXXXXXXXXXXXXXXXXXXXXXXXXXXXXXXXXXXXXXXXXXXXXXXXXXXXXXXXXXX
%XXXXXXXXXXXXXXXXXXXXXXXXXXXXXXXXXXXXXXXXXXXXXXXXXXXXXXXXXXXXXXXXXXXXXXXX
%XXXXXXXXXXXXXXXXXXXXXXXXXXXXXXXXXXXXXXXXXXXXXXXXXXXXXXXXXXXXXXXXXXXXXXXXX

\chapter{CONDITIONS FOR QUASI-EXACT SOLVABILITY}

\section{Introduction to QES}

In this chapter we study QES model in one dimension.  These are
the models for which a part of the bound state energy spectrum and
corresponding wave-functions can be obtained exactly.  These
models have been constructed and studied extensively by means of
Lie algebraic approach.  For a review we refer to the book by
Ushveridze et al [6].  In order that a part of the spectrum be
obtained exactly, the potential parameters appearing in the
potential must satisfy a condition known as the condition for
quasi-exact solvability.  Within the QHJ approach, as used for
exactly solvable models, it is not clear how such a condition can
arise and why only a part of the spectrum is exactly solvable.  In
this chapter we report a study of these aspects of QES models.

In order to study QES models within QHJ formalism one needs to
have information of singularities of QMF.  This in general is not
very easy to obtain except for some simple cases like harmonic
oscillator and hydrogen atom problems.  In the limit $\hbar
\longrightarrow 0 $ the QMF $p(x,E)$ goes over to
$p_{cl}(x,E)=\sqrt{E-V(x)} $ which will, in general, have several
branch points.  This is an indication that in general, the
singularity structure of $p(x,E) $ will be very complicated. {\it{
\bf{In order to make progress, we make a simplifying assumption
that the point at infinity is an isolated singular point and more
specifically it is a pole of some finite order.}}}  Thus this
amounts to saying that $ p(x,E) $ has fixed poles, and a finite
number of moving poles and a pole at infinity.  Using these
requirements, we can proceed as in the case of exactly solvable
models and work out the consequences of exact quantization
condition given below.
\begin{equation}
\oint{p}dq = nh.     \label{4.1a}
\end{equation}

We find that for all the QES potential models studied by us,
(\ref{4.1a}) and our assumptions, imply that potential parameter
must satisfy a condition which turns out to be identical with the
condition of quasi-exact solvability of the potential. A list of
potentials studied and the condition of quasi-exact solvability in
each case are given below.

The potentials are:
\begin{enumerate}
\item Sextic oscillator:
      \begin{equation}
      V(x)=\alpha x^{2}+\beta x^{4}+\gamma x^{6}, \qquad
      \gamma>0.
      \end{equation}
\item Sextic oscillator with centrifugal barrier:
      \begin{equation}
      V(x)=4(s-\frac{1}{4})(s-\frac{3}{4})\frac{1}{x^{2}}
      +[b^{2}-4a(s+\frac{1}{2}+\mu)]x^{2}+2abx^{4}+a^{2}x^{6}.
      \end{equation}
\item Circular potential:
      \begin{equation}
      V(x)=\frac{A}{\sin^{2}x}+\frac{B}{\cos ^{2}x} +C\sin ^{2}x -D\sin
      ^{4}x,
      \end{equation}
with
      \begin{equation}
      A=4(s_{1}-\frac{1}{4})(s_{1}-\frac{3}{4}),
      \end{equation}
      \begin{equation}
      B=4(s_{2}-\frac{1}{4})(s_{2}-\frac{3}{4}),
      \end{equation}
      \begin{equation}
      C=q_{1}^{2}+4q_{1}(s_{1}+s_{2}+\mu),
      \end{equation}
      \begin{equation}
      D=q_{1}^{2}.
      \end{equation}
\item Hyperbolic potential:
      \begin{equation}
      V(x) = -\frac{A}{\cosh^{2}x} + \frac{B}{\sinh^{2}x}
      -C \cosh^{2}x + D\cosh^{4}x,
      \end{equation}
      with
      \begin{equation}
      A=4(s_{1}-\frac{1}{4})(s_{1}-\frac{3}{4}),
      \end{equation}
      \begin{equation}
      B= 4(s_{2}- \frac{1}{4})(s_{2}-\frac{3}{4}),
      \end{equation}
      \begin{equation}
      C = [q_{1}^{2}+4q_{1}(s_{1}+s_{2}+\mu)],
      \end{equation}
      \begin{equation}
      D=  q_{1}^{2}.
      \end{equation}

\item \begin{equation}V(x)=A\sinh^{2}\sqrt{\nu}x+B\sinh
      {\sqrt{\nu}}x+C\tanh{\sqrt{\nu}}x \sech{\sqrt{\nu}}x+
      D\sech^{2}{\sqrt{\nu}}x
      \end{equation}
\item \begin{equation}V(x)=A\cosh^{2}\sqrt{\nu}x+B\cosh
      {\sqrt{\nu}}x+C\coth{\sqrt{\nu}}x\ \csc h{\sqrt{\nu}}x+D
          \csch^{2}{\sqrt{\nu}}x
          \end{equation}
    \item \begin{equation}V(x)= Ae^{2\sqrt{\nu} x} +Be^{\sqrt{\nu} x}+Ce^{-\sqrt{\nu}
          x}+De^{-2\sqrt{\nu} x}
          \end{equation}
    \end{enumerate}
    The conditions for quasi exact solvability for these potentials
    are:
    \begin{enumerate}
    \item
          \begin{equation}
          \frac{1}{\sqrt{\gamma}}\left(\frac{\beta^{2}}{4\gamma}
          -\alpha \right)=3+2n, \qquad n=\mbox{\rm integer}
          \end{equation}
    \item
          \begin{equation}
          \mu = \mbox{\rm integer}
          \end{equation}
    \item  Taking
    \begin{equation}
          A=4(s_{1}-\frac{1}{4})(s_{1}-\frac{3}{4})
          \end{equation}
          \begin{equation}
          B=4(s_{2}-\frac{1}{4})(s_{2}-\frac{3}{4})
          \end{equation}
          \begin{equation}
          C=q_{1}^{2}+4q_{1}(s_{1}+s_{2}+\mu)
          \end{equation}
          \begin{equation}
          D=q_{1}^{2}
          \end{equation}
          the condition is
          \begin{equation}
          \mu = \mbox{\rm integer}
          \end{equation}
    \item Taking
          \begin{equation}
          A=4(s_{1}-\frac{1}{4})(s_{1}-\frac{3}{4}),
          \end{equation}
          \begin{equation}
          B= 4(s_{2}- \frac{1}{4})(s_{2}-\frac{3}{4}),
          \end{equation}
          \begin{equation}
          C = [q_{1}^{2}+4q_{1}(s_{1}+s_{2}+\mu)],
          \end{equation}
          \begin{equation}
          D=  q_{1}^{2},
          \end{equation}
          the condition is
          \begin{equation}
          \mu = \mbox{\rm integer}
          \end{equation}
    \item
          \begin{equation}\left[ B\pm 2(n+1)\sqrt{\nu A} \right]^{4}+A(4D-\nu )\left[B \pm
        2(n+1)\sqrt{\nu A}\right]^{2} -4A^{2}C^{2}=0
          \end{equation}
    \item
          \begin{equation}\left[ B\pm 2(n+1)\sqrt{\nu A} \right]^{4}-A(4D+\nu )\left[B \pm
          2(n+1)\sqrt{\nu A}\right]^{2} +4A^{2}C^{2}=0
          \end{equation}
    \item
          \begin{equation} 2(n+1)\sqrt{\nu AD} = \pm B\sqrt{D} \pm C\sqrt{A}
          \end{equation}
    \end{enumerate}

    The calculations for these potentials are given in the next few
    sections.

    %XXXXXXXXXXXXXXXXXXXXXXXXXXXXXXXXXXXXXXXXXXXXXXXXXXXXXXXXXXXXXXXXXXXXXXXXXX
    %XXXXXXXXXXXXXXXXXXXXXXXXXXXXXXXXXXXXXXXXXXXXXXXXXXXXXXXXXXXXXXXXXXXXXXXXXX
    \section{A Representation of QES Quantization Rule}

    We now bring out some common features of the exactly solvable
    models, that have been studied in this thesis and those reported in the
    paper [5].   For the exactly solvable model the QMF written in
    terms  of suitable variables $y$ takes the form
    \begin{equation}
    p(y)=\frac{b_{1}}{y-\xi_{1}}+\frac{b_{1}^{\prime}}{y-\xi_{2}}+\cdots
    +\sum_{k=1}^{n}\frac{-i}{y-y_{k}} +R(y)
    \end{equation}
    where $\xi_{1}, \xi_{2} ,\cdots  $ are fixed poles, the summation
    term corresponds to $n$ moving poles at $y_{1}, y_{2},\cdots ,
    y_{n}$ and $R(y)$ is atmost a polynomial in $y$.  The residues
    $b_{1}, b_{1}^{\prime},\cdots $  have been calculated using the
    QHJ equation and demanding a condition such as the one proposed by
    Leacock and Padgett, or

    \[ \lim_{E \to 0} p(x,E) = i\ W(x), \]
     or the
    square integrability of the wave-function.  Thus in all the cases
    studied we are lead to a rational expression for the QMF.

    Under an assumption about the behavior of QMF at infinity, even
    for QES models, the QMF turns out to be a rational function.  The
    quantization rule
    \begin{equation}
    \oint {p(x,E)}dx =nh
    \end{equation}
    is then easily seen to be equivalent to the well know result, that
    for a rational function, sum of residues at all poles, including
    the one at infinity vanishes.  Written explicitly for a rational
    form of QMF that we have, this requirement becomes
    \begin{equation}
    \sum_{\rm fixed \ poles} \left(\mbox{\rm Res of  QMF} \right)
        + n + \left(
    \mbox{\rm Res of QMF  at  infinity} \right) = 0 \label{4.2a}
    \end{equation}
    where Res stand for the residue and the middle term $n$,
    corresponds to the contribution of moving poles to the residue.

    In this and the next chapter, we will use this condition
    (\ref{4.2a}) as a substitute for quantization rule.

    %XXXXXXXXXXXXXXXXXXXXXXXXXXXXXXXXXXXXXXXXXXXXXXXXXXXXXXXXXXXXXXXXXXXXXX
    %XXXXXXXXXXXXXXXXXXXXXXXXXXXXXXXXXXXXXXXXXXXXXXXXXXXXXXXXXXXXXXXXXXXXXX
    \section{Sextic Oscillator}

    The potential for the sextic oscillator is:
    \begin{equation}
    V(x)=\alpha x^{2}+\beta x^{4}+\gamma x^{6}, \qquad \gamma>0
    \end{equation}

    The QHJ equation is ($\hbar=1=2m$)
    \begin{equation}
    p^{2}(x,E) -ip^{\prime}(x,E)-(E-V)=0   \label{4.3a}
    \end{equation}

    For the $ n^{th} $ excited state, the QMF has n poles on the real
    axis and we assume that there are no other moving poles. We shall
    use the quantization condition viz.,
    \begin{equation}
    \frac{1}{2\pi}\oint_{C}{p(x,E)dx}=n\hbar  \label{4.3b}
    \end{equation}
    in the form (\ref{4.2a}) as given above.

    To evaluate the integral in (\ref{4.3b})  a Laurent expansion of
    $\tilde{p}(y)$ in powers of $y=1/x$, is made
    \begin{equation}
    \tilde{p}(y)=\frac{b_{3}}{y^{3}}+\frac{b_{2}}{y^{2}}+\frac{b_{1}}{y}
    +a_{0}+a_{1}y+\cdots   \label{4.3c}
    \end{equation}
    Substituting this in (\ref{4.3a}) and integrating term by term we
    get
    \begin{equation}
    J(E)=ia_{1}
    \end{equation}
    The quantization condition gives
    \begin{equation}
    a_{1}=-in
    \end{equation}
    It only remains to compute the coefficient $a_{1}$ of the Laurent
    expansion given in (\ref{4.3c}). To do this we start from the QHJ
    equation
    \begin{equation}
    \tilde{p}^{2}(y)
    +iy^{2}\tilde{p}^{\prime}(y)-E+\frac{\alpha}{y^{2}}
    +\frac{\beta}{y^{4}}+\frac{\gamma}{y^{6}}=0
    \end{equation}
    Substituting the Laurent expansion and equating the coefficients
    of different powers of $y$ on both sides of the equation we get
    \begin{equation}
    b_{3}=\pm i\sqrt{\gamma}
    \end{equation}
    \begin{equation}
    b_{1}=-\frac{\beta}{2}b_{3}   \label{4.3d}
    \end{equation}
    \begin{equation}
    a_{1}=\frac{(-\alpha -b_{1}^{2}+3ib_{3})}{2b_{3}} \label{4.3e}
    \end{equation}

    It is important to know that, we would get two solutions for $
    b_{1}$ corresponding to the two solutions of $b_{3}=\pm
    i\sqrt{\gamma}. $  This happens due to the fact that the QHJ is
    quadratic in the QMF.  Thus one needs a boundary condition to pick
    the correct solution.  We propose to use the square integrability
    of the wave-function instead of the original boundary condition,
    explained in chapter 2, which was proposed by Leacock and Padgett.
    This is because the original boundary condition is difficult to
    implement in the present case due to the presence of six branch
    points in the $p_{cl}. $  In order to find the restrictions coming
    from the square  integrability,  we compute the wave-function
    \begin{equation}
    \psi(x) = \exp\left(\int {ip(x)}dx\right)
    \end{equation}
    for large $x$ as follows.  The most important term in the Laurent
    expansion (\ref{4.3c}) for small $ y \approx 0$, corresponding to
    large $x$ is
    \begin{equation}
    \tilde{p}(y) \approx \frac{b_{3}}{y^{3}}
    \end{equation}
    and the wave-function for large $x$ becomes
    \begin{equation}
    \psi (x) \approx \exp\left(i\frac{b_{3}x^{4}}{4}\right)
    \end{equation}
    Out of the two solutions, $b_{3}=\pm i\sqrt{\gamma}, \  \psi(x)$
    is square integrable only for $b_{3}= i\sqrt{\gamma}$.  Using this
    value of $b_{3}$ and from (\ref{4.3d}) and  (\ref{4.3e}),
     equating $a_{1}$ to $ -in $ we get
    \begin{equation}
    \frac{1}{\sqrt{\gamma}}\left(\frac{\beta ^{2}}{4\gamma} -\alpha
    \right) = 3 + 2n  \label{4.3f}
    \end{equation}
    In order to compare the results in (\ref{4.3f}) with the well
    known condition, we write
    \begin{equation}
    \gamma = a^{2}, \ \beta=2ab
    \end{equation}
    Thus we get $ \alpha = b^{2}-a(3+2n) $ which agree with the result
    given in [6].

    %XXXXXXXXXXXXXXXXXXXXXXXXXXXXXXXXXXXXXXXXXXXXXXXXXXXXXXXXXXXXXXXXXXXXXXXXX
    %XXXXXXXXXXXXXXXXXXXXXXXXXXXXXXXXXXXXXXXXXXXXXXXXXXXXXXXXXXXXXXXXXXXXXXXXX

    \section{Sextic Oscillator with a Centrifugal Barrier}

    The potential is given as
    \begin{equation}
    V(x)=4(s-\frac{1}{4})(s-\frac{3}{4})\frac{1}{x^{2}}
    +[b^{2}-4a(s+\frac{1}{2}+\mu)]x^{2}+2abx^{4}+a^{2}x^{6}
    \end{equation}
    We shall consider only the case $s>\frac{3}{4}$ so that the coefficient
    of the centrifugal term, $\frac{1}{x^{2}}$ is positive.
    The Q.H.J equation is ($\hbar=1=2m$)
    \begin{equation}
    p^{2}(x,E) -ip^{\prime}(x,E)-(E-V)=0
    \end{equation}
    Substituting the potential  the QHJ equation is
    \begin{equation}
    p^{2}(x,E)
    -ip^{\prime}(x,E)-[E-4(s-\frac{1}{4})(s-\frac{3}{4})\frac{1}{x^{2}}
    -[b^{2}-4a(s+\frac{1}{2}+\mu)]x^{2}-2abx^{4}-a^{2}x^{6}=0
    \label{4.4a}
    \end{equation}
    $ p(x,E) $ has poles at $x=0$  and as the potential is symmetric
    there are moving poles on either side of the origin.   We assume
    that there are no more poles in the complex plane.  We assume that
    infinity is a pole. We find below the residues for each of these
    pole.

    We expand $p(x,E)$ as
    \begin{equation}
    p(x,E)=\frac{b{_1}}{x}+a_{0}+a_{1}x+\cdots  \label{4.4b}
    \end{equation}
    Using (\ref{4.4b}) in (\ref{4.4a})  and equating the coefficient of
    $\frac{1}{x^{2}}$ we get
    \begin{equation}
    b_{1}=-\frac{i}{2}\left[1\pm  (4s-2)\right]
    \end{equation}
    Demanding that the wave-function remains finite, for $x
         \rightarrow 0$, gives

    \begin{equation}
    b_{1}=-\frac{i}{2}\left[4s-1)\right]
    \end{equation}

    To find residue for the pole at infinity, we use the mapping $
    x=\frac{1}{t}$. Therefore equation (\ref{4.4a}) transforms to

    \[
    \tilde{p}^{2}(t,E) +it^{2}\tilde{p}^{\prime}(t,E)
    \]
    \begin{equation}
    -\left[E-4(s-\frac{1}{4})(s-\frac{3}{4})t^{2}
    -[b^{2}-4a(s+\frac{1}{2}+\mu)]\frac{1}{t^{2}}-2ab\frac{1}{t^{4}}
    -a^{2}\frac{1}{t^{6}}\right]=0  \label{4.4c}
    \end{equation}

    We expand $\tilde{p}(t,E)$  as
    \begin{equation}
    \tilde{p}(t,E) =
    \frac{d_{3}}{t^{3}}+\frac{d_{2}}{t^{2}}+\frac{d_{1}}{t}
    +c_{0}+c_{1}t+c_{2}t^{2}+\cdots  \label{4.4d}
    \end{equation}

    Using (\ref{4.4d}) in (\ref{4.4c}) and equating different  coefficient
    of  $t$ we get

    \begin{equation}
    d_{3}= \pm ia
    \end{equation}

    \begin{equation}
    d_{2}=\frac{-ab}{d_{3}}
    \end{equation}
    \begin{equation}
     c_{0}=0
    \end{equation}
    \begin{equation}
    c_{1}=\frac{2a(s+\frac{1}{2}+\mu)}{d_{3}}+\frac{3i}{2}
    \end{equation}
    The correct sign of $d_{3}$ is fixed by the condition of square
    integrability and is given by
    \begin{equation}
    d_{3} = -  ia
    \end{equation}
    Now equating the sum of all residues to zero, we get
    \begin{equation}
    ib_{1} +2n =c_{1}
    \end{equation}
    Substituting the values of $b_{1} $and $c_{1}
     $ in the above relation
    yields the required condition, viz
    \begin{equation}
    n=\mu
    \end{equation}
    The above condition agrees with those given in [6]

    %XXXXXXXXXXXXXXXXXXXXXXXXXXXXXXXXXXXXXXXXXXXXXXXXXXXXXXXXXXXXXXXXXXXXXXXX
    %XXXXXXXXXXXXXXXXXXXXXXXXXXXXXXXXXXXXXXXXXXXXXXXXXXXXXXXXXXXXXXXXXXXXXXXXXX

    \section{Circular Potential}

    The potential is given as
    \begin{equation}
    V(x)=\frac{A}{\sin^{2}x}+\frac{B}{\cos ^{2}x} +C\sin ^{2}x -D\sin
    ^{4}x
    \end{equation}
    where
    \begin{equation}
    A=4(s_{1}-\frac{1}{4})(s_{1}-\frac{3}{4})
    \end{equation}
    \begin{equation}
    B=4(s_{2}-\frac{1}{4})(s_{2}-\frac{3}{4})
    \end{equation}
    \begin{equation}
    C=q_{1}^{2}+4q_{1}(s_{1}+s_{2}+\mu)
    \end{equation}
    \begin{equation}
    D=q_{1}^{2}
    \end{equation}
    We effect a change of variable by
    \begin{equation}
    y=\sin ^{2} x
    \end{equation}
    The Q.H.J equation is
    \begin{equation}
    p^{2}(x,E) -ip^{\prime}(x,E)-(E-V)=0
    \end{equation}
    In the new variable the QHJ equation is
    \begin{equation}
    \tilde{p}^{2}(y,E) -2i\sqrt{y}\sqrt{1-y}
    \tilde{p}^{\prime}(y,E)-\left[E-\frac{A}{y} -\frac{B}{1-y}
    -Cy+Dy^{2}\right] =0
    \end{equation}
    Let
    \begin{equation}
    \tilde{p}(y,E)=-2i\sqrt{y}\sqrt{1-y}\phi
    \end{equation}
    In terms of the above transformation the QHJ becomes
    \begin{equation}
    \phi ^{2}+\phi^{\prime} +\frac{1}{2}\frac{(1-2y)}{y(1-y)}\phi
    +\frac{1}{4}\frac{1}{y(1-y)}\left[E-\frac{A}{y} -\frac{B}{1-y}
    -Cy+Dy^{2}\right] =0  \label{4.5a}
    \end{equation}
    $ \phi $ has poles at $y=0$ and at $ y=+ 1 $ \ and there are a
    finite number of moving poles in the complex plane.  We assume
    that there are no more poles in the complex plane. We find the
    residues for each of these pole below.

    For  $ y=0 $  we consider an expansion in  $\phi $  as
    \begin{equation}
    \phi = \left(  \frac{b_{1}}{y} + a_{0} + a_{1}y + \cdots \right)
    \label{4.5b}
    \end{equation}
    Using (\ref{4.5b}) in (\ref{4.5a}) and equating the various powers
    of $y$ we get the following.  The power of $ \frac{1}{y^{2}}$
    gives
    \begin{equation}
    b_{1} = \frac{1}{2}\left[ \frac{1}{2} \pm (2s_{1}-1) \right]
    \end{equation}
    The correct value of $b_{1}$ is fixed by the condition of square
    integrability of the wave-function and is given below as
    \begin{equation}
    b_{1} = \frac{1}{2}\left[ \frac{1}{2} + (2s_{1}-1) \right]
    \end{equation}

    For $ y=1 $ we consider an expansion in $\phi $ as
    \begin{equation}
    \phi= \left(  \frac{b_{1}^{\prime}}{y-1} + a_{0}^{\prime} +
    a_{1}^{\prime}(y-1) + \cdots \right)  \label{4.5c}
    \end{equation}
    Using (\ref{4.5c}) in (\ref{4.5a}) and equating the various powers
    of $y$ we get the following.  The power of $ \frac{1}{(y-1)^{2}}$
    gives
    \begin{equation}
    b_{1}^{\prime} = \frac{1}{2}\left[ \frac{1}{2} \pm (2s_{2}-1)
    \right]
    \end{equation}

    The correct value of $b_{1}^{\prime}$ is fixed by the condition of
    square integrability of the wave-function and is given below as
    \begin{equation}
    b_{1}^{\prime} = \frac{1}{2}\left[ \frac{1}{2} + (2s_{2}-1)
    \right]
    \end{equation}

    To find residue for the pole at infinity, we use the mapping $
    y=\frac{1}{t}$. Therefore equation (\ref{4.5a}) transforms to

    \[
    \tilde{\phi}^{2}(t,E) -t^{2}\tilde{\phi}^{\prime}(t,E)
    +\frac{1}{2}\frac{t(t-2)}{t-1}\tilde{\phi} (t,E)
    \]
    \begin{equation}
    \left[ \frac{E}{4}\frac{t^{2}}{t-1}-\frac{A}{4}\frac{t^{3}}{t-1}
    -\frac{B}{4}\frac{t^{3}}{(t-1)^{2}}+\frac{C}{4}\frac{t}{t-1}
    +\frac{D}{4}\frac{1}{t-1} \right] \label{4.5d}
    \end{equation}

    We expand $\tilde{\phi}(t,E)$  as
    \begin{equation}
    \tilde{\phi}(t,E) = \frac{d_{1}}{t}+c_{0}+c_{1}t+\cdots
    \label{4.5e}
    \end{equation}

    Using (\ref{4.5e}) in (\ref{4.5d}) and equating the  power of $
    \frac{1}{t^{2}} $ we get

    \begin{equation}
    d_{1}= 0
    \end{equation}
    Equating the constant term we have
    \begin{equation}
    c_{0} = \pm \frac{q_{1}}{2}
    \end{equation}
    Equating the coefficient  of $\frac{1}{t} $ gives the residue at 
    $y=\infty$ as
    
    \begin{equation}
    c_{1}= \frac{q_{1}(s_{1}+s_{2}+\mu)}{2c_{0}} - \frac{1}{2}
    \end{equation}

    The correct sign of $c_{0}$ is fixed by the condition of square
    integrability and is given by
    \begin{equation}
    c_{0} =  \frac{q_{1}}{2}
    \end{equation}

    Now equating the sum of all residues of fixed poles and the moving
    poles and the pole at infinity, we have the following relation.
    \begin{equation}
    b_{1}+b_{1}^{\prime}+n=c_{1}
    \end{equation}
    Substituting the values of $b_{1},b_{1}^{\prime},
     $ and $ c_{1}$ in the above relation
    yields the required condition
    \begin{equation}
    \mu = n
    \end{equation}
    The above condition agrees with those given in [6]

    %XXXXXXXXXXXXXXXXXXXXXXXXXXXXXXXXXXXXXXXXXXXXXXXXXXXXXXXXXXXXXXXXXXXXXXXXX
    %XXXXXXXXXXXXXXXXXXXXXXXXXXXXXXXXXXXXXXXXXXXXXXXXXXXXXXXXXXXXXXXXXXXXXXXXXX

    \section{Hyperbolic Potential}

    The  hyperbolic potential is
    \begin{equation}
       V(x) =
    -\frac{A}{\cosh^{2}x} + \frac{B}{\sinh^{2}x}
     -C \cosh^{2}x + D\cosh^{4}x
    \end{equation}
    where
    \[A=4(s_{1}-\frac{1}{4})(s_{1}-\frac{3}{4}) \]
    \[B= 4(s_{2}- \frac{1}{4})(s_{2}-\frac{3}{4})
    \]
     \[ C = [q_{1}^{2}+4q_{1}(s_{1}+s_{2}+\mu)]  \]
    \[ D=  q_{1}^{2} \]

     We will consider the case $s_{2}>\frac{3}{4}$.

     The Q.H.J equation is ($ \hbar =1=2m  $)
     \begin{equation}
    p^{2}(x,E) -i p^{\prime}(x,E) -(E-V) =0
     \end{equation}
    We use a mapping is $ y = \cosh x $

    The Q.H.J equation in the new variable is:

    \[ \tilde{p}(y,E) -i\hbar\sqrt{y^{2}-1}p^{\prime}(y,E) - [E+
    \frac{A}{y^{2}}-\frac{B}{y^{2}-1}+Cy^{2}-Dy^{4}]=0 \]

    Let \begin{equation}
     \tilde{p}(y,E) = -i \sqrt{y^{2}-1}\phi(y,E)
    \end{equation}
    and hence
    \begin{equation}
     \tilde{p} \ ^  {\prime}(y,E) = -i \{ \sqrt{y^{2}-1}\phi^{\prime}
    +\phi\frac{y}{\sqrt{y^{2}-1}} \}
    \end{equation}
    Therefore the Q.H.J. equation becomes
    \begin{equation}
    \phi^{2}+\phi^{\prime}+\frac{y}{y^{2}-1}\phi+\frac{1}{y^{2}-1}[E+
    \frac{A}{y^{2}}-\frac{B}{y^{2}-1}+Cy^{2}-Dy^{4}]=0
    \end{equation}
    
    We note that  $\phi$ has  fixed poles  at $y=\pm 1$ and $y=0$ and we write
    the Q.H.J equation  as
    \[
    \phi^{2}+\phi^{\prime}+[\frac{1}{2}\frac{1}{y+1}+\frac{1}{2}\frac{1}{y-1}]
    \phi+E[\frac{1}{2}\frac{1}{y-1}-\frac{1}{2}\frac{1}{y+1}]+A[-\frac{1}{y^{2}}-\frac{1}{2}\frac{1}{y+1}+
    \frac{1}{2}\frac{1}{y-1}]  \]
    \[ -B[-\frac{1}{4}\frac{1}{y-1}+\frac{1}{4}\frac{1}{(y-1)^{2}}+\frac{1}{4}
    \frac{1}{y+1}+\frac{1}{4}\frac{1}{(y+1)^{2}}]  \]
    \begin{equation}
     +C[1+ \frac{1}{2}\frac{1}{y-1}-\frac{1}{2}\frac{1}{y+1}]
    -D[1+y^{2}+\frac{1}{2}\frac{1}{y-1}-\frac{1}{2}\frac{1}{y+1}]=0
    \label{4.6a}
    \end{equation}

    For  \ $ y=0 $

    Let \[ \phi = \frac{b_{1}}{y} + \sum_{n=0}^{\infty}a_{n}y^{n} \]

    \[ -\frac{i}{2\pi}\oint{\phi}dy =
    -\frac{i}{2\pi}\oint{(\frac{b_{1}}{y}+
    \sum_{n=0}^{\infty}a_{n}y^{n})}dy  =-\frac{i}{2\pi}2\pi ib_{1}
    =b_{1} \]

    The QHJ equation (\ref{4.6a}) becomes

    \[ (\frac{b_{1}}{y} +a_{0}+a_{1}y+\ldots)^{2}
    +(-\frac{b_{1}}{y^{2}}+a_{1}+\ldots)+
    [\frac{1}{2}\frac{1}{y+1}+\frac{1}{2}\frac{1}{y-1}]
    \phi+E[\frac{1}{2}\frac{1}{y-1}-\frac{1}{2}\frac{1}{y+1}] \]
    \[ +A[-\frac{1}{y^{2}}-\frac{1}{2}\frac{1}{y+1}+
    \frac{1}{2}\frac{1}{y-1}]
    -B[-\frac{1}{4}\frac{1}{y-1}+\frac{1}{4}\frac{1}{(y-1)^{2}}+\frac{1}{4}
    \frac{1}{y+1}+\frac{1}{4}\frac{1}{(y+1)^{2}}] \]
    \begin{equation}
      +C[1+
    \frac{1}{2}\frac{1}{y-1}-\frac{1}{2}\frac{1}{y+1}]
    -D[1+y^{2}+\frac{1}{2}\frac{1}{y-1}-\frac{1}{2}\frac{1}{y+1}]=0
    \end{equation}

    The coefficient of $ \frac{1}{y^{2}} $ gives
    \begin{equation}
    b_{1}^{2}-b_{1}-A=0
    \end{equation}
    and hence
    \begin{equation}
    b_{1} =\frac{1}{2} \pm(2s_{1}-1)
    \end{equation}

    For  $ y=1 $ , let
    \begin{equation}
     \phi = \frac{b^{\prime}_{1}}{y-1} +
    \sum_{n=0}^{\infty}a^{\prime}_{n}(y-1)^{n} \label{4.6b}
    \end{equation}
    Therefore
    \[ -\frac{i}{2\pi}\oint{\phi}dy = -\frac{i}{2\pi}\oint{[\frac{b^{\prime}_{1}}{(y-1)}
    + \sum_{n=0}^{\infty}a^{\prime}_{n}(y-1)^{n}]}dy
    =-\frac{i}{2\pi}2\pi ib^{\prime}_{1} =b^{\prime}_{1} \]

    Substituting (\ref{4.6b}) in (\ref{4.6a}) and equating the
    coefficient of $\frac{1}{(y-1)^{2}}$ we get

    \begin{equation}
     {b^{\prime}_{1}}^{2}
    -\frac{1}{2}b^{\prime}_{1}-B\frac{1}{4} =0
    \end{equation}
    which yields

    \begin{equation}
    b^{\prime}_{1} = \frac{1}{4} \pm \frac{1}{2}(2s_{2}-1)
    \end{equation}

    For  $ y=-1 $, let

    \begin{equation}
     \phi = \frac{b^{{\prime}{\prime}}_{1}}{y+1} +
    \sum_{n=0}^{\infty}a^{{\prime}{\prime}}_{n}(y+1)^{n} \label{4.6c}
    \end{equation}
    Therefore
    \[ -\frac{i}{2\pi}\oint{\phi}dy = -\frac{i}{2\pi}\oint{[\frac{b^{{\prime}{\prime}}_{1}}{(y+1)}
    + \sum_{n=0}^{\infty}a^{{\prime}{\prime}}_{n}(y+1)^{n}]}dy
    =-\frac{i}{2\pi}2\pi ib^{{\prime}{\prime}}_{1}
    =b^{{\prime}{\prime}}_{1}
    \]

    Substituting (\ref{4.6c}) in (\ref{4.6a}) and equating the
    coefficient of $ \frac{1}{(y+1)^{2}} $ we get
    \begin{equation}
    {b^{{\prime}{\prime}}_{1}}^{2}
    -\frac{1}{2}b^{{\prime}{\prime}}_{1}-\frac{B}{4} =0
    \end{equation}
    On simplification yields the same value as $ b^{\prime}_{1}  $.
    Hence
    \begin{equation}
    b^{{\prime}{\prime}}_{1} = \frac{1}{4} \pm \frac{1}{2}(2s_{2}-1)
    \end{equation}
    To calculate the residue at $ y=\infty   $, we apply the mapping $
    y=\frac{1}{u}  $

    The QHJ equation (\ref{4.6a}) becomes
    \[
    \tilde{\phi}^{2}(u,E)+(-u^{2})\tilde{\phi}^{\prime}(u,E)+[\frac{1}{2}\frac{u}{1+u}+\frac{1}{2}\frac{u}{1-u}]\tilde{\phi}(u,E)
    \]
    \[+E[\frac{1}{2}\frac{u}{1+u}
    -\frac{1}{2}\frac{u}{1+u}]+A[-u^{2}-\frac{1}{2}\frac{u}{1+u}+\frac{1}{2}\frac{u}{1-u}]
    \]
    \[  -B[-\frac{1}{4}\frac{u}{1-u}+
    \frac{1}{4}\frac{u^{2}}{(1-u^){2}}+\frac{1}{4}\frac{u}{1+u}+\frac{1}{4}\frac{u^{2}}{(1+u)^{2}}]+C[1+\frac{1}{2}\frac{u}{1-u}-\frac{1}{2}\frac{u}{1+u}]
    \]
    \begin{equation}
     -D[1+\frac{1}{u^{2}}+\frac{1}{2}\frac{u}{1-u}-\frac{1}{2}\frac{u}{1+u}]=0
     \end{equation}
    Let
    \begin{equation}
    \tilde{\phi}(u,E)=\frac{c_{1}}{u}+d_{0}+d_{1}u+\cdots
    \end{equation}
    The residue at infinity is the coefficient of $d_{1}$ and is given
    as
    \begin{equation}
    d_{1}=-\frac{2q_{1}(s_{1}+s_{2}+\mu)}{c_{1}}-1
    \end{equation}
    with
    \begin{equation}
    c_{1}=\pm q_{1}
    \end{equation}

    Equating the sum of all residues of fixed poles, the moving
    poles to the pole at infinity, we have the following relation.
    \begin{equation}
    d_{1}=2n+b_{1}+b_{1}^{\prime}+b_{1}^{{\prime}{\prime}}
    \end{equation}
    The equation (4.6.6) does not change when replacements
    $y \rightarrow -y$ and $\phi \rightarrow - \phi$ is made.
    Therefore we select the residue $b_{1}^{\prime} = b_{1}^{{\prime}{\prime}}$
    and the condition of finiteness of the wave-function at $ x=0$ requires that
    positive sign be selected in (4.6.12) and (4.6.15).  Therefore

\begin{equation}
b_{1}^{\prime}=b_{1}^{{\prime}{\prime}}=s_{2}-\frac{1}{4}
\end{equation}
The point $y=0$ corresponds to complex value of $x$ and therefore
one cannot insist on finiteness of the wave-function at $x=0$. One
must fall back on the boundary condition given in chapter 2
section 2.3.1.  We will simply note that selecting positive sign
in  (4.4.19) leads us to the correct condition

    \begin{equation}
    \mu=n
    \end{equation}
for quasi exact solvability [6]. In this chapter, our objective
has been to show that QES conditions follows from our assumption
that point at inifnity is an isolated singular point.  In the
cases where for certain ranges of potential parameters, both the
residues are acceptable, one must
 accept all such answers and work out the consequences.  This may lead to
 some new and interesting results as is evidenced by the investigations on phases of
super symmetry [13] and periodic potentials [15].

 Besides the above
 QES potentials, we  now take up three classes of QES potentials [9,10]
 and find the conditions for quasi exact solvability within our approach.

    %XXXXXXXXXXXXXXXXXXXXXXXXXXXXXXXXXXXXXXXXXXXXXXXXXXXXXXXXXXXXXXXXXXXXXXXXXX
    %XXXXXXXXXXXXXXXXXXXXXXXXXXXXXXXXXXXXXXXXXXXXXXXXXXXXXXXXXXXXXXXXXXXXXXXXXX
    \section{$V(x)=A\sinh^{2}\sqrt{\nu}x+B\sinh
    {\sqrt{\nu}}x+C\tanh{\sqrt{\nu}}x\ \sech{\sqrt{\nu}}x+
           D\sech^{2}{\sqrt{\nu}}x $}
    The potential is given as
    \begin{equation}
    V(x)=A\sinh^{2}\sqrt{\nu}x+B\sinh
    {\sqrt{\nu}}x+C\tanh{\sqrt{\nu}}x \sech{\sqrt{\nu}}x+D\sec
    h^{2}{\sqrt{\nu}}x
    \end{equation}
    We effect a change of variable by
    \begin{equation}
    y=\sinh \sqrt{\nu}x
    \end{equation}
    The QHJ equation is  $(\hbar=1=2m)$
    \begin{equation}
    p^{2}(x,E) -ip^{\prime}(x,E)-(E-V)=0
    \end{equation}
    In the new variable the QHJ equation is
    \begin{equation}
    \tilde{p}^{2}(y,E) -i\sqrt{\nu}\sqrt{1+y^{2}}
    \tilde{p}^{\prime}(y,E)-\left[E-Ay^{2} -By -C\frac{y}{1+y^{2}}
    -D\frac{1}{1+y^{2}}\right] =0
    \end{equation}

    Let
    \begin{equation}
    \tilde{p}(y,E)=-i\sqrt{\nu}\sqrt{1+y^{2}}\phi
    \end{equation}
    In terms of the above transformation the QHJ becomes
    \[
    [\phi+\frac{1}{2}\frac{y}{1+y^{2}}]^{2}+\phi^{\prime}
    -\frac{1}{4}\frac{y^{2}}{(1+y^{2})^{2}}
    \]
    \begin{equation}
    +\frac{1}{\nu (1+y^{2})}\left[E-Ay^{2} -By -C\frac{y}{1+y^{2}}
    -D\frac{1}{1+y^{2}}\right] =0
    \end{equation}

    Let
    \begin{equation}
    \chi=\phi+\frac{1}{2}\frac{y}{1+y^{2}}
    \end{equation}
    Therefore  the above equation  becomes
    \[
    \chi^{2}+\chi^{\prime}+\frac{3}{4}\frac{y^{2}}{(1+y^{2})^{2}}
    -\frac{1}{2}\frac{1}{1+y^{2}}
    \]
    \begin{equation}
    +\frac{1}{\nu (1+y^{2})}\left[E-Ay^{2} -By -C\frac{y}{1+y^{2}}
    -D\frac{1}{1+y^{2}}\right] =0  \label{4.7a}
    \end{equation}
    $ \chi $ has poles at  $ y=\pm i $ \ and there are moving poles
    between the turning points.  We assume that there are no more
    poles in the complex line. We find the residues for each of these
    pole below.

    For$ y=i $ we consider an expansion in $\chi $ as
    \begin{equation}
    \chi = \left(  \frac{b_{1}}{y-i} + a_{0} + a_{1}(y-i) + \cdots
    \right) \label{4.7b}
    \end{equation}
    Using (\ref{4.7b}) in (\ref{4.7a}) and equating the various powers
    of $y$ we get the following.  The power of $ \frac{1}{(y-i)^{2}}$
    gives
    \begin{equation}
    b_{1} = \frac{1}{2}\left[ 1 \pm \frac{1}{2}
    \sqrt{1-4(\frac{D}{\nu}+\frac{iC}{\nu})} \right]
    \end{equation}

    For$ y=-i $  we consider an expansion in $\chi $ as
    \begin{equation}
    \chi = \left(  \frac{b_{1}^{\prime}}{y+i} + a_{0}^{\prime} +
    a_{1}^{\prime}(y+i) + \cdots \right) \label{4.7c}
    \end{equation}
    Using (\ref{4.7c}) in (\ref{4.7a}) and equating the  power of $
    \frac{1}{(y+i)^{2}}$ gives
    \begin{equation}
    b_{1}^{\prime} = \frac{1}{2}\left[ 1 \pm \frac{1}{2}
    \sqrt{1-4(\frac{D}{\nu}-\frac{iC}{\nu})} \right]
    \end{equation}

    To find the pole at infinity, we use the mapping $ y=\frac{1}{t}$
    Therefore equation (\ref{4.7a}) transforms to

    \[
    \chi^{2}(t,E) -t^{2}\chi^{\prime}(t,E)
    +\frac{3}{4}\frac{t^{2}}{(t^{2}+1)^{2}} -
    \frac{1}{2}\frac{t^{2}}{t^{2}+1}
    \]
    \begin{equation}
    +\frac{1}{\nu (t^{2}+1)}[Et^{2}-A-Bt-C\frac{t^{5}}{t^{2}+1}
    -D\frac{t^{6}}{t^{2}+1}]=0 \label{4.7d}
    \end{equation}
    For the point at infinity we expand $\chi$  as
    \begin{equation}
    {\chi}(t,E) = d_{0}+d_{1}t+d_{2}t^{2}+\cdots \label{4.7e}
    \end{equation}
    Using (\ref{4.7e}) in (\ref{4.7d}) and equating the constant term
    we get

    \begin{equation}
    d_{0}= \pm \sqrt{\frac{A}{\nu}}
    \end{equation}
    and equating the  term  in $t $ we have
    \begin{equation}
    d_{1} = \frac{B}{2\nu d_{0}}
    \end{equation}

    Now equating the sum of residues due to fixed poles  moving poles
    and that at infinity,to zero we have the following relation.
    \begin{equation}
    b_{1}+b_{1}^{\prime}+n=d_{1}
    \end{equation}
    Substituting the values of $b_{1},b_{1}^{\prime} $ and $ d_{1}$ in
    the above relation yields the required condition given below

    \[ B^{4}+16A^{2}\nu^{2}(n+1)^{4}
     +24AB^{2}\nu (n+1)^{2}  +4AB^{2}D-AB^{2}\nu
    (n+1)^{2}
    \]
    \[
    \pm 32AB\nu \sqrt{A\nu }(n+1)^{3} \pm 4AB\nu \sqrt{A\nu}(n+1)
    -AB^{2}\nu- 4A^{2}\nu^{2}(n+1)^{2}
    \]
    \begin{equation}
    +16A^{2}D\nu (n+1)^{2}\pm 16ABD\sqrt{A\nu}(n+1) \pm
    8B^{3}\sqrt{A\nu}(n+1) - 4A^{2}C^{2}
    \end{equation}
    This can be written in the compact form as:
    \begin{equation}
    \left[ B\pm 2(n+1)\sqrt{\nu A} \right]^{4}+A(4D-\nu )\left[B \pm
    2(n+1)\sqrt{\nu A}\right]^{2} -4A^{2}C^{2}=0
    \end{equation}
    The above condition agrees with those given in [9,10]

    %XXXXXXXXXXXXXXXXXXXXXXXXXXXXXXXXXXXXXXXXXXXXXXXXXXXXXXXXXXXXXXXXXXXXXXXXXXX
    %XXXXXXXXXXXXXXXXXXXXXXXXXXXXXXXXXXXXXXXXXXXXXXXXXXXXXXXXXXXXXXXXXXXXXXXXXXXX

    \section{$V(x)=A\cosh^{2}\sqrt{\nu}x+B\cosh
          {\sqrt{\nu}}x+C\coth{\sqrt{\nu}}x \csch{\sqrt{\nu}}x+D
              \csch^{2}{\sqrt{\nu}}x $ }

The potential is given as
\begin{equation}
V(x)=A\cosh^{2}\sqrt{\nu}x+B\cosh
{\sqrt{\nu}}x+C\coth{\sqrt{\nu}}x \csch{\sqrt{\nu}}x+D
\csch^{2}{\sqrt{\nu}}x
\end{equation}
We effect a change of variable by
\begin{equation}
y=e^{\sqrt{\nu}x}
\end{equation}
The QHJ equation is $(\hbar=1=2m)$
\begin{equation}
p^{2}(x,E) -ip^{\prime}(x,E)-(E-V)=0
\end{equation}
In the new variable the QHJ equation is
\[
\tilde{p}^{2}(y,E) -i\sqrt{\nu} \tilde{p}^{\prime}(y,E)
\]
\begin{equation}
-\left[E-A\frac{1}{4y^{2}}(y^{2}+1)^{2}
-B\frac{1}{2y}{(y^{2}+1)^{2}} -C\frac{2y(y^{2}+1)}{(y^{2}-1)^{2}}
-D\frac{4y^{2}}{(y^{2}-1)^{2}}\right] =0
\end{equation}
Let
\begin{equation}
\tilde{p}(y,E)=-i\sqrt{\nu}y\phi
\end{equation}
In terms of the above transformation the QHJ becomes
\[
[\phi +\frac{1}{2y}]^{2}+\phi^{\prime} -\frac{1}{4y^{2}}
\]
\begin{equation}
+\frac{1}{\nu y^{2}}\left[E-A\frac{1}{4y^{2}}(y^{2}+1)^{2}
-B\frac{1}{2y}{(y^{2}+1)^{2}} -C\frac{2y(y^{2}+1)}{(y^{2}-1)^{2}}
-D\frac{4y^{2}}{(y^{2}-1)^{2}}\right] =0
\end{equation}
Let
\begin{equation}
\chi=\phi+\frac{1}{2y}
\end{equation}
Therefore the above  equation  becomes
\[
\chi^{2}+\chi^{\prime}+\frac{1}{4y^{2}}
\]
\begin{equation}
+\frac{1}{\nu y^{2}}\left[E-A\frac{1}{4y^{2}}(y^{2}+1)^{2}
-B\frac{1}{2y}{(y^{2}+1)^{2}} -C\frac{2y(y^{2}+1)}{(y^{2}-1)^{2}}
-D\frac{4y^{2}}{(y^{2}-1)^{2}}\right] =0 \label{4.8a}
\end{equation}
$ \chi $ has poles at $y=0$ and at $ y=\pm 1 $ \ and there are
moving poles between the turning points.  We assume that there are
no more poles in the complex line. We find the residues for each
of these pole below.

For  $ y=0 $  we consider an expansion in  $\chi $  as
\begin{equation}
\chi = \left( \frac {b_{2}}{y^{2}}+ \frac{b_{1}}{y} + a_{0} +
a_{1}y + \cdots \right)  \label{4.8b}
\end{equation}
Using (\ref{4.8b}) in (\ref{4.8a}) and equating the power of $
\frac{1}{y^{4}}$ gives
\begin{equation}
b_{2} = \pm \sqrt{\frac{A}{4\nu}}
\end{equation}
The power of $ \frac{1}{y^{3}}$ gives
\begin{equation}
b_{1}=1+\frac{B}{4\nu b_{2}}
\end{equation}
For $y=1$  we consider an expansion in $\chi$ as
\begin{equation}
\chi = \left(  \frac{b_{1}^{\prime}}{y-1} + a_{0}^{\prime} +
a_{1}^{\prime}(y-1) + \cdots \right)  \label{4.8c}
\end{equation}
Using (\ref{4.8c}) in (\ref{4.8a}) and equating the  power of $
\frac{1}{(y-1)^{2}}$ gives
\begin{equation}
b_{1}^{\prime} = \frac{1}{2}\left[ 1 \pm
\sqrt{1+4(\frac{D}{\nu}+\frac{C}{\nu})} \right]
\end{equation}
For  $y=-1$  we consider an expansion in $\chi $ as
\begin{equation}
\chi = \left(  \frac{b_{1}^{{\prime}{\prime}}}{y+1} +
a_{0}^{{\prime}{\prime}} + a_{1}^{{\prime}{\prime}}(y+1) + \cdots
\right)  \label{4.8d}
\end{equation}
Using (\ref{4.8d}) in (\ref{4.8a}) and equating the  power of $
\frac{1}{(y+1)^{2}}$ gives
\begin{equation}
b_{1}^{{\prime}{\prime}} = \frac{1}{2}\left[ 1 \pm
\sqrt{1+4(\frac{D}{\nu}-\frac{C}{\nu})} \right]
\end{equation}
To find the pole at infinity, we use the mapping $ y=\frac{1}{t}$.
Therefore equation (\ref{4.8a}) transforms to

\[
\tilde{\chi}^{2}(t,E) -t^{2}\tilde{\chi}^{\prime}(t,E)
+\frac{1}{4}t^{2}
\]
\begin{equation}
+\frac{1}{\nu}t^{2}
[E-\frac{A}{4}\frac{(1+t^{2})^{2}}{t^{2}}-\frac{B}{2}\frac{1+t^{2}}{t}
-2C\frac{(1+t^{2})t}{(1-t^{2})^{2}}-4D\frac{t^{2}}{(1-t^{2})^{2}}]=0
\label{4.8e}
\end{equation}

we expand $\tilde{\chi}(t,E)$ as
\begin{equation}
\tilde{\chi}(t,E) = d_{0}+d_{1}t+d_{2}t^{2}+\cdots \label{4.8f}
\end{equation}
Using (\ref{4.8f}) in (\ref{4.8e}) and equating the constant term
we get

\begin{equation}
d_{0}= \pm \sqrt{\frac{A}{4\nu}}
\end{equation}
Equating the  term  in $t $ we have
\begin{equation}
d_{1} = \frac{B}{4\nu d_{0}}
\end{equation}
Now equating the sum of residues due to fixed poles, the moving
poles, and that at infinity, to zero we have the following
relation.
\begin{equation}
b_{1}+b_{1}^{\prime}+b_{1}^{{\prime}{\prime}}+2n=d_{1}
\end{equation}
Substituting the values of $b_{1},b_{1}^{\prime},
b_{1}^{{\prime}{\prime}} $ and $ d_{1}$ in the above relation
yields the required condition This can be written in the compact
form as:
\begin{equation}
\left[ B\pm 2(n+1)\sqrt{\nu A} \right]^{4}-A(4D+\nu )\left[B \pm
2(n+1)\sqrt{\nu A}\right]^{2} +4A^{2}C^{2}=0
\end{equation}
The above condition agrees with those given in [9,10]

%XXXXXXXXXXXXXXXXXXXXXXXXXXXXXXXXXXXXXXXXXXXXXXXXXXXXXXXXXXXXXXXXXXXXXXXXXX
%XXXXXXXXXXXXXXXXXXXXXXXXXXXXXXXXXXXXXXXXXXXXXXXXXXXXXXXXXXXXXXXXXXXXXXXXXXX

\section{$V(x)= Ae^{2\sqrt{\nu} x} +Be^{\sqrt{\nu} x}+Ce^{-\sqrt{\nu}
      x}+De^{-2\sqrt{\nu} x} $ }

The potential is
\begin{equation}
V(x)= Ae^{2\sqrt{\nu} x} +Be^{\sqrt{\nu} x}+Ce^{-\sqrt{\nu}
x}+De^{-2\sqrt{\nu} x}
\end{equation}

The QHJ equation is given by $(\hbar=1=2m)$
\begin{equation}
p^{2}(x,E) -ip^{\prime}(x,E) -[E-Ae^{2\sqrt{\nu} x}
-Be^{\sqrt{\nu} x}-Ce^{-\sqrt{\nu} x}-De^{-2\sqrt{\nu} x}=0
\end{equation}

We use a change of variable by
\begin{equation}
y=e^{\sqrt{\nu}x}
\end{equation}

Therefore the QHJ transforms to

\begin{equation}
\tilde{p}^{2}(y,E)-i\sqrt{\nu}y\tilde{p}^{\prime}(y,E)-[E-Ay^{2}-By-C\frac{1}{y}
-D\frac{1}{y^{2}}]=0
\end{equation}

Let
\begin{equation}
\tilde{p}(y,E) =-i\sqrt{\nu}y\phi
\end{equation}
Therefore the above equation becomes

\begin{equation}
(\phi + \frac{1}{2y})^{2}+\phi ^{\prime}
-\frac{1}{4y^{2}}+\frac{1}{\nu
y^{2}}\left[E-Ay^{2}-By-C\frac{1}{y} -D\frac{1}{y^{2}}\right]=0
\end{equation}
Let

\begin{equation}
\chi = \phi + \frac{1}{2y}
\end{equation}

With this transformation we get

\begin{equation}
\chi ^{2} +\chi ^{\prime} + \frac{1}{4y^{2}}+\frac{1}{\nu
y^{2}}\left[E-Ay^{2}-By-C\frac{1}{y} -D\frac{1}{y^{2}}\right]=0
\label{4.9a}
\end{equation}

$ \chi $ has poles at $ y=0 $ and there are moving poles between
the turning points.  We assume that there are no more poles in the
complex line other than a pole at infinity.  We compute the
residue for $y=0$

For  $ y=0 $ we define

\begin{equation}
\chi = \frac{b_{2}}{y^{2}}+\frac{b_{1}}{y}+a_{0}+a_{1}y+\cdots
\label{4.9b}
\end{equation}

Substituting (\ref{4.9b}) in (\ref{4.9a}) we get
\[
( \frac{b_{2}}{y^{2}}+\frac{b_{1}}{y}+a_{0}+a_{1}y+\cdots)^{2}
+(-2\frac{b_{2}}{y^{3}}-\frac{b_{1}}{y^{2}}+a_{1}+\cdots)+
\frac{1}{4y^{2}}
\]
\begin{equation}
+\frac{1}{\nu y^{2}}\left[E-Ay^{2}-By-C\frac{1}{y}
-D\frac{1}{y^{2}}\right]=0
\end{equation}

Equating the coefficient of $ \frac{1}{y^{4}} $we get
\begin{equation}
b_{2}= \pm \sqrt{\frac{D}{\nu}}
\end{equation}

Equating the coefficient of $ \frac{1}{y^{3}} $we get

\begin{equation}
b_{1} = 1+\frac{C}{2b_{2}\nu}
\end{equation}

We assume that infinity is a pole and compute the residue at
infinity, we which we use the mapping

\begin{equation}
y= \frac{1}{t}
\end{equation}

Therefore (\ref{4.9a}) transforms to
\begin{equation}
\tilde{\chi}^{2}(t,E) - t^{2}\tilde{\chi}^{\prime}(t,E)
+\frac{t^{2}}{4}+\frac{t^{2}}{\nu}\left[E-A\frac{1}{t^{2}}-B\frac{1}{t}-Ct-Dt^{2}\right]
=0   \label{4.9c}
\end{equation}
We use a Laurent's  expansion of the form

\begin{equation}
\tilde{\chi}(t,E)=d_{0}+d_{1}t +d_{2}t^{2}+\cdots  \label{4.9d}
\end{equation}

Using (\ref{4.9d}) in (\ref{4.9c}) and equating the of $ t $ the
constant term yields

\begin{equation}
d_{0}= \pm \sqrt{\frac{A}{\nu}}
\end{equation}
and the coefficient of $ t $ yields
\begin{equation}
d_{1}= \frac{B}{2d_{0}\nu}
\end{equation}

Now equating the sum of residues due to fixed poles, the moving
poles, and that at infinity, to zero we have the following
relation.
\begin{equation}
b_{1}+n = d_{1}
\end{equation}
Substituting the values of $ b_{1}$ and $ d_{1} $ in the above
equation we get.
\begin{equation}
\pm \frac{C\sqrt{\nu}}{2\nu\sqrt{D}}+(n+1) = \pm
\frac{B\sqrt{\nu}}{2\nu\sqrt{A}}
\end{equation}
The above on simplification gives the desired condition given
below which agrees with that given in [9,10].
\begin{equation}
2(n+1)\sqrt{\nu AD} = \pm B\sqrt{D} \pm C\sqrt{A}
\end{equation}

%XXXXXXXXXXXXXXXXXXXXXXXXXXXXXXXXXXXXXXXXXXXXXXXXXXXXXXXXXXXXXXXXXXXXXXXXX
%XXXXXXXXXXXXXXXXXXXXXXXXXXXXXXXXXXXXXXXXXXXXXXXXXXXXXXXXXXXXXXXXXXXXXXXXX

\section{Quartic Oscillator}

We end this chapter with a short analysis of quartic an-harmonic
oscillator and give some remarks on polynomial potentials of
degree different from six.

First we consider the $x^{4}$ oscillator with
\begin{equation}
V(x)= \alpha x + \beta x^{2} + \gamma x^{3} + \delta x^{4}, \qquad
\delta > 0
\end{equation}
We ask whether this model is QES for any choice of parameters.  We
repeat the analysis given for sextic oscillator, assuming that the
point at infinity is an isolated singular point, a pole of some
order $m$. We therefore substitute
\begin{equation}
p(x,E)=b_{m}x^{m}+b_{m-1}x^{m-1}+\cdots
\end{equation}
in the QHJ equation and determine the constants $b_{m}$. 
 For $m >2$  we find that  $b_{m}=0$ and $b_{2}=\pm i\sqrt{\delta}$.  
Thus corresponding bound state wave-function for large $x$ will
behave as
\begin{equation}
\psi(x) \sim \exp\left(i\frac{b_{2}}{3}x^{3}\right)
\end{equation}
and for both the choices $\pm i\sqrt{\delta}$ for $b_{2}$, one
gets wave-function which grows either at $+$ infinity or $-$
infinity. Hence the assumptions, that the point at infinity is an
isolated singular point of QMF, is inconsistent with QHJ for real
parameter $\alpha, \beta, \gamma$ and $\delta$.  Thus, $x^{4}$
oscillator does not lead to any choice of real parameter.  However
for complex parameters, one can get the known results for QES
quartic model  [18].

%XXXXXXXXXXXXXXXXXXXXXXXXXXXXXXXXXXXXXXXXXXXXXXXXXXXXXXXXXXXXXXXXXXXXXXXXXXXXX
%XXXXXXXXXXXXXXXXXXXXXXXXXXXXXXXXXXXXXXXXXXXXXXXXXXXXXXXXXXXXXXXXXXXXXXXXXXXXX
\section{Summary and Observations}

From our study in this chapter we arrive at the following
conclusions.
\begin{enumerate}
\item For the QES models, the QMF corresponding to the
      algebraic part of the spectrum has singularity structure
      very similar to the exactly solvable models.
\item The integer $n$ appearing in the exact quantization condition
      is just the number of moving poles of QMF in the complex
      plane.  In the case of exactly solvable models the moving
      poles are in a one to one correspondence with the  real nodes
      of the wave-functions, but a corresponding statement for QES
      model is not true.  This result and some other interesting
      properties will be explicitly demonstrated in the next
      chapter for bound state wave-functions of QES models.
\item The integer $n$ in the right hand side of the quantization
      condition appears in the condition of quasi-exact
      solvability of the potential.
\item The condition of quasi-exact solvability is equivalent to our
      assumption about the behavior of QMF at infinity, reflecting
      the simplification of the singularity structure for the QES
      bound states.
\end{enumerate}

%XXXXXXXXXXXXXXXXXXXXXXXXXXXXXXXXXXXXXXXXXXXXXXXXXXXXXXXXXXXXXXXXXXXXXXXXXXXXXXXXX
%XXXXXXXXXXXXXXXXXXXXXXXXXXXXXXXXXXXXXXXXXXXXXXXXXXXXXXXXXXXXXXXXXXXXXXXXXXXXXXXXX
%XXXXXXXXXXXXXXXXXXXXXXXXXXXXXXXXXXXXXXXXXXXXXXXXXXXXXXXXXXXXXXXXXXXXXXXXXXXXXXXXX

\chapter {CALCULATION OF WAVE-FUNCTIONS FOR QES MODELS}

In the previous chapter we have seen that the condition for quasi-exact
solvability arises from a simple requirement on the behavior of
QMF at infinity. We continue our study of QES models and  take
up an investigation of the  wave-functions.  We find that the
wave-functions can be computed by proceeding as in the case of
exactly solvable models.  We begin with our simplifying assumption
mentioned in the previous chapter for the QES models and proceed
in the same fashion as for the case of exactly solvable models in
chapter 3.  Thus the QMF is meromorphic and the corresponding
residues at the poles are known, and also the behavior at infinity
is known, with this information the bound state wave-functions can
be obtained as in chapter 3.  We give our results for two
potential models viz. the sextic oscillator and the hyperbolic
potential. This study reveals a new interesting feature of the
zeros of the wave-functions, which will be discussed at the end of
this chapter.

%XXXXXXXXXXXXXXXXXXXXXXXXXXXXXXXXXXXXXXXXXXXXXXXXXXXXXXXXXXXXXXXXXXXXXXXXXXX
%XXXXXXXXXXXXXXXXXXXXXXXXXXXXXXXXXXXXXXXXXXXXXXXXXXXXXXXXXXXXXXXXXXXXXXXXXXXXXXX

\section{Sextic Oscillator}

The potential for the sextic oscillator is:
\begin{equation}
V(x) =\alpha x^{2}+\beta x^{4} +\gamma x^{6}, \qquad \gamma > 0
\end{equation}
with the following values for $\alpha ,\beta ,\gamma $ and the
condition for $\mu, n, p$ where  $p$ stands for parity

$ \alpha = b^{2} - a(3+2n) , \  \beta=2ab  , \ \gamma = a^{2} , \
4\mu + 2p = 2n  \ $with  $\ p=0  \ or  \ 1  $

 The QHJ equation is ($\hbar=1=2m$)
\begin{equation}
p(x,E) - i p^{\prime}(x,E) - (E - \alpha x^{2}-\beta x^{4} -\gamma
x^{6}) = 0  \label{5.1a}
\end{equation}
We assume that the point at infinity is a pole.  Therefore $p(x,E)
$ behaves as $x^{n}$ for some $n$

\[ p(x,E) \sim x^{n}
\]
for  large  $x$. Hence $p(x,E)$ takes the form for large $x$.
\begin{equation}
   p(x,E) = a_{3}x^{3} + a_{2}x^{2} + a_{1}x +
   a_{0}+O(\frac{1}{x})  \label{5.1b}
\end{equation}
where $a_{0},a_{1}\cdots , a_{3}$ are constants, on the assumption
that $p(x,E)$ have no other singular points and substitute
(\ref{5.1b})  in equation (\ref{5.1a}).  Next we equate the
coefficient of powers of $x^{6}$ to zero, gives
\begin{equation}
a_{3}^{2} +\gamma =0
\end{equation}
Since $\gamma = a^{2}$, we have
\begin{equation}
 a_{3} = \pm ia
\end{equation}

As  $ a_{3} $ has two values, the correct value is fixed by the
condition of square integrability on the wave function.

\[ \psi(x) = \exp\left(i\int{p(x,E)}dx\right)
= \exp \left(i\int{(a_{3}x^{3} + a_{1}x)}dx \right) \]
 If the
above integral have to bounded at infinity, the we require that
\begin{equation}
a_{3} = + ia \end{equation}

Next equating the coefficient of successive powers
$x^{5},x^{4},\ldots $ to zero we get
\begin{equation}
 a_{2} =0
\end{equation}
\begin{equation}
 a_{1} = - \frac{ab}{a_{3}}
\end{equation}
\begin{equation}
  a_{0} = 0
\end{equation}
Hence
\begin{equation}
 a_{1} =  ib
\end{equation}

Therefore p(x,E) becomes
\begin{equation}
 p(x,E) =\sum_{k=1}^{n} \frac{-i}{x-x_{k}} + iax^{3} + ibx +c
 \label{5.1c}
 \end{equation}
To determine $x_{k}$, or equivalently $
P(x)=\prod_{k=1}^{n}(x-x_{k}),$ we substitute (\ref{5.1c}) in
(\ref{5.1a}) and get

\[ \left(-i\frac{P^{\prime}(x)}{P(x)} + iax^{3} +ibx +c\right)^{2}
+ \frac{P^{{\prime}{\prime}}(x)}{P(x)} - (\frac{
P^{\prime}(x)}{P(x)})^{2}+ 3ax^{2} + b
\]
\begin{equation}
- [ E - \alpha x^{2} -\beta x^{4} -\gamma x^{6}] = 0
\end{equation}
Therefore, the above equation becomes
\begin{equation}
 c^{2} + 2ax^{3}\frac{P^{\prime}}{P} + 2ibcx
-2i\frac{P^{\prime}}{P}c + 2bx\frac{P^{\prime}}{P} + 2iacx^{3} -
\frac{P^{{\prime}{\prime}}}{P} + b - E - 2anx^{2} = 0
\end{equation}
Equating the coefficient of  $x^{3} $  term we have
\[ 2iacx^{3} =0 \Longrightarrow c = 0 \]
and hence we have
\begin{equation}
  2ax^{3}\frac{P^{\prime}}{P} + 2bx\frac{P^{\prime}}{P}
- \frac{P^{{\prime}{\prime}}}{P} + b - E - 2anx^{2} = 0
\label{5.1d}
\end{equation}
The above equation thus gives the following differential equation
in $P(x)$
\begin{equation}
P^{{\prime}{\prime}} - P^{\prime}(2ax^{3} + 2bx) - P(b- E
-2anx^{2}) =0 \label{5.1e}
\end{equation}
We get the expression for energies and wave functions  for various
values of $ n $ as follows:

 We will derive explicit form of
wave-functions for $n=0,1$ and $2$.  Later we will discuss the
general form of the wave-function for arbitrary $n$.  The general
strategy for obtaining the wave-functions is the same as discussed
for exactly solvable models in chapter 3.

{\it\bf{ Wave-function for $n$=0}}:  Only one energy level can be
solved in this case.  Since the number $n$, representing the
number of moving poles is zero (\ref{5.1c}), with $c$=0 as already
found, becomes
\begin{equation}
p(x,E)=iax^{3}+ibx
\end{equation}
and hence the wave-function is given by
\begin{equation}
 \psi(x) = \exp\left({i\int{p(x)dx}}\right) = \exp\left({i\int[ iax^{3} +
 ibx]dx}\right)=\exp\left({-a\frac{x^{4}}{4} -
 b\frac{x^{2}}{2}}\right)
\end{equation}
and the corresponding energy is obtained from (\ref{5.1e}) by
equating the constant term and is given as
\begin{equation}
E=b.
\end{equation}

{\it\bf{ Wave-function for $n$=1}}:  In this case we take $P(x,E)$
to be a first degree polynomial, $(x-x_{0})$.  There (\ref{5.1e})
gives, $x_{0}=0$  and the energy is given as
\begin{equation}
E=3b.
\end{equation}
Therefore the wave-function comes out to be
\begin{equation}
\psi(x)=Nx\exp\left({-a\frac{x^{4}}{4} - b\frac{x^{2}}{2}}\right).
\end{equation}

{\it \bf{Wave-function for $n$=2}}:  We seek a solution of
(\ref{5.1e}) with $P(x)$ as a second degree polynomial.
Substituting $P(x)$ as
\begin{equation}
P(x) = \alpha_{0} +\alpha_{1}x + \alpha_{2}x^{2}.
\end{equation}
Using the above equation in (\ref{5.1e}) and comparing different
powers of $x$ gives
\begin{equation}
\alpha_{1}=0
\end{equation}
\begin{equation}
  4a\alpha_{0}
-\alpha_{2}(5b - E) =0 \label{5.1a1}
\end{equation}
\begin{equation}
\alpha_{0}(b-E)-2\alpha_{2}  = 0.  \label{5.1a2}
\end{equation}

The last two equation have non-trivial solution for $\alpha_{0}$
and $\alpha_{2}$ only if the
\[ \left |
    \begin{array}{cc}
    4a     & (5b-E) \\
    (b-E) &  -2
    \end{array}
    \right | =0.
\]
This gives two energy eigen-values

\begin{equation}
E = 3b \pm  2\sqrt{b^{2} + 2a}.
\end{equation}
To get the wave function we compute $\alpha_{1}$ and$\alpha_{2}$
from equation (\ref{5.1a1}) and (\ref{5.1a2}) and use
\begin{equation}
p(x) = -i\frac{2\alpha_{2} x}{\alpha_{0} + \alpha_{2}x^{2}} +
iax^{3} + ibx.
\end{equation}

Therefore the wave function is given by:
\begin{equation}
\psi(x) =N \exp\left({i\int{p(x)dx}}\right) =N
\exp\left({i\int[-i\frac{2\alpha_{2} x}{\alpha_{0} +
\alpha_{2}x^{2}} + iax^{3} + ibx]dx}\right).
\end{equation}
\begin{equation}
 \psi(x) =N (\alpha_{0} + \alpha_{2}x^{2}) \exp\left({-a\frac{x^{4}}{4} - b\frac{x^{2}}{2}}\right)
\end{equation}
where $N$ is the normalizing factor. The value of $\alpha_{0}$ is
given by
\begin{equation}
\alpha_{0}=\frac{5b-E}{4a}.
\end{equation}
Replacing the value of $\alpha_{0}$ and energy value $E$ in the
above equation one gets the expression for wave-function as

\begin{equation}
\psi(x)= N\frac{E-5b}{4a}\left[ b \pm \sqrt{b^{2}+2a}  x^{2} -1
\right]\exp\left({-a\frac{x^{4}}{4} - b\frac{x^{2}}{2}}\right).
\end{equation}

The wave-functions and eigen-values explicitly obtained for the
cases $n=0,1$ and 2 agree with the known results [6].

For an arbitrary value of $n$ the polynomial $P(x)$ will be
obtained by solving (\ref{5.1e}).  If we take $P(x)$ to be of the
form
\begin{equation}
P(x)=\sum_{k=0}^{} \alpha_{k}x^{k},
\end{equation}
then the differential equation (\ref{5.1e}) leads to a set of
homogenous equations for the corresponding coefficients
$\alpha_{0},\alpha_{1},\cdots,\alpha_{n}$.  These equations will
have a non-trivial solution only if determinant of the
coefficients vanishes. This condition will determine the energy
eigen-value, corresponding to each eigen-value we can find the
coefficients $\alpha_{0},\alpha_{1},\cdots,\alpha_{n}$.  Thus we
get $n$ independent wave-function each having the form
\begin{equation}
\psi(x) \sim P(x)\exp\left({-a\frac{x^{4}}{4} -
b\frac{x^{2}}{2}}\right).
\end{equation}
Notice that all these eigen-functions corresponding to a fixed
value of $n$ have a olynomial of the same degree $n$ as a factor.
Thus for a fixed value of $n$, and hence for a given set of
potential parameters, wave-functions for all the states which can
be solved have the same number of zeros equal to $n$.  If these
levels are arranged according to increasing energy, the number of
zeros on the real axis (nodes) will increase.  Hence the number of
complex zero will decrease with increasing energy.  This feature
appears to be a general property of quasi-exactly solvable models.

%XXXXXXXXXXXXXXXXXXXXXXXXXXXXXXXXXXXXXXXXXXXXXXXXXXXXXXXXXXXXXXXXXXXXXXXXXX
%XXXXXXXXXXXXXXXXXXXXXXXXXXXXXXXXXXXXXXXXXXXXXXXXXXXXXXXXXXXXXXXXXXXXXXXXXXXXXX

\section{Hyperbolic Potential}

The hyperbolic potential is
\begin{equation}
   V(x) = -\frac{A}{\cosh^{2}x} + \frac{B}{\sinh^{2}x}
 -C \cosh^{2}x + D\cosh^{4}x,
\end{equation}
where
\[A=4(s_{1}-\frac{1}{4})(s_{1}-\frac{3}{4}), \]
\[B= 4(s_{2}- \frac{1}{4})(s_{2}-\frac{3}{4}),
\]
 \[ C = [q_{1}^{2}+4q_{1}(s_{1}+s_{2}+\mu)],  \]
\[ D=  q_{1}^{2}. \]
The QHJ equation is  ($\hbar=1=2m$)
\begin{equation}
p^{2}(x,E) -i p^{\prime}(x,E) -[E+\frac{A}{\cosh^{2}x} -
\frac{B}{\sinh^{2}x}
 +C \cosh^{2}x - D\cosh^{4}x] =0.
 \end{equation}
We effect a transformation by
\begin{equation}
  y = \cosh x
\end{equation}

The QHJ equation in the new variable is
\begin{equation}
 \tilde{p}(y,E) -i\hbar\sqrt{y^{2}-1}p^{\prime}(y,E) - [E+
\frac{A}{y^{2}}-\frac{B}{y^{2}-1}+Cy^{2}-Dy^{4}]=0
\end{equation}
Let
\begin{equation}
\tilde{p}(y,E) = -i \sqrt{y^{2}-1}\phi(y,E)
\end{equation}
Therefore the QHJ  equation becomes
\begin{equation}
 \left[ \phi+\frac{1}{2}\frac{y}{y^{2}-1}\right]^{2}-\frac{1}{4}\frac{y^{2}}
{(y^{2}-1)^{2}} +\phi ^{\prime} + \frac{1}{y^{2}-1}[E+
\frac{A}{y^{2}}-\frac{B}{y^{2}-1}+Cy^{2}-Dy^{4}]=0.
\end{equation}
Let
\begin{equation}
\chi=\phi+\frac{1}{2}\frac{y}{y^{2}-1}.
\end{equation}
Therefore the above equation becomes
\begin{equation}
\chi^{2}+\chi^{\prime}+\frac{3}{4}\frac{y^{2}}{(y^{2}-1)^{2}}-\frac{1}{2}\frac{1}{y^{2}-1}
+ \frac{1}{y^{2}-1}[E+
\frac{A}{y^{2}}-\frac{B}{y^{2}-1}+Cy^{2}-Dy^{4}]=0  \label{5.2a}
\end{equation}
$ \chi$ has poles at $ y=0,$ and \ $y= \pm 1$, and there are
moving poles between the turning points.  We assume that there are
no more poles in the complex plane other than a pole at infinity.
We will first compute the residues at $y=0, \pm1$ and then in the
general form of $\chi$ (\ref{5.2i}) the constants
$b_{1},b_{1}^{\prime}$ and $b_{1}^{{\prime}{\prime}}$ will be
known and then we give the general form of the wave-function.

{\it{\bf {Computation of residues}}}: For $y=0$ let
\begin{equation}
\chi =\frac{b_{1}}{y}+a_{0}+a_{1}y+\cdots
\end{equation}
Therefore equation (\ref{5.2a}) becomes

\[
[\frac{b_{1}}{y}+a_{0}+a_{1}y+\cdots]^{2}+[-\frac{b_{1}}{y^{2}}+a_{1}+\cdots]
+\frac{3}{4}\frac{y^{2}}{(y^{2}-1)^{2}}-\frac{1}{2}\frac{1}{y^{2}-1}
\]
\begin{equation}
+ \frac{1}{y^{2}-1}[E+
\frac{A}{y^{2}}-\frac{B}{y^{2}-1}+Cy^{2}-Dy^{4}]=0
\end{equation}
Equating the coefficient of $\frac{1}{y^{2}}$ on both sides gives
\begin{equation}
b_{1} =\frac{1}{2}[1 \pm (4s_{1}-2)]
\end{equation}
By the condition of square integrability, the positive sign has to
be taken.  Hence the value of $b_{1}$ is
\begin{equation}
 b_{1} = 2s_{1}-\frac{1}{2}
\end{equation}
For $y=1$ let
\begin{equation}
\chi
=\frac{b_{1}^{\prime}}{y-1}+a_{0}^{\prime}+a_{1}^{\prime}(y-1)+\cdots
\end{equation}
Therefore equation (\ref{5.2a}) becomes

\[
[\frac{b_{1}^{\prime}}{y-1}+a_{0}^{\prime}+a_{1}^{\prime}(y-1)+\cdots]^{2}+[-\frac{b_{1}^{\prime}}{(y-1)^{2}}+a_{1}^{\prime}+\cdots]
+\frac{3}{4}\frac{y^{2}}{(y^{2}-1)^{2}}-\frac{1}{2}\frac{1}{y^{2}-1}\]
\[ + \frac{1}{y^{2}-1}[E+
\frac{A}{y^{2}}-\frac{B}{y^{2}-1}+Cy^{2}-Dy^{4}]=0
\]
Equating the coefficient of $\frac{1}{(y^-1){2}} $on both sides
gives
\begin{equation}
b_{1}^{\prime}=\frac{1}{2}[1 \pm (2s_{2}-1) \ ]
\end{equation}
By the condition of square integrability, the positive sign has to
be taken.  Hence the value of $b_{1}^{\prime}$ is
\begin{equation}
b_{1}^{\prime} = s_{2}
\end{equation}
For $y=-1$  let
\begin{equation}
\chi
=\frac{b_{1}^{{\prime}{\prime}}}{y+1}+a_{0}^{{\prime}{\prime}}+a_{1}^{{\prime}{\prime}}(y+1)+\cdots
\end{equation}
Therefore equation (\ref{5.2a}) becomes

\[
[\frac{b_{1}^{{\prime}{\prime}}}{y+1}+a_{0}^{{\prime}{\prime}}
+a_{1}^{{\prime}{\prime}}(y+1)+\cdots]^{2}+[-\frac{b_{1}^{{\prime}{\prime}}}{(y+1)^{2}}+a_{1}^{{\prime}{\prime}}+\cdots]
+\frac{3}{4}\frac{y^{2}}{(y^{2}-1)^{2}}-\frac{1}{2}\frac{1}{y^{2}-1}\]
\[ + \frac{1}{y^{2}-1}[E+
\frac{A}{y^{2}}-\frac{B}{y^{2}-1}+Cy^{2}-Dy^{4}]=0
\]
Equating the coefficient of$\frac{1}{(y^+1){2}}$ on both sides
gives
\begin{equation}
b_{1}^{{\prime}{\prime}}=\frac{1}{2}[1 \pm (2s_{2}-1) \ ]
\end{equation}
By the condition of square integrability, the positive sign has to
be taken.  Hence the value of $b_{1}^{{\prime}{\prime}}$ is
\begin{equation}
b_{1}^{{\prime}{\prime}}=s_{2}.
\end{equation}
For the fixed poles at $y=0,\pm 1 $ let $\chi$ have the form
\begin{equation}
\chi
=\frac{P^{\prime}(y)}{P(y)}+\frac{b_{1}}{y}+\frac{b_{1}^{\prime}}{y-1}
+\frac{b_{1}^{{\prime}{\prime}}}{y+1}+c_{1}y+c_{2},  \label{5.2i}
\end{equation}
where $c_{1}$ and $c_{2}$ are constants to be determined.  This
form of $\chi$ is because the equation (\ref{5.2a}) has a $y^{2}$
term.

{\it{\bf{Form of wave-function}}}: Using (\ref{5.2i}), equation
(\ref{5.2a}) transforms to
\[ \left[ \frac{P^{\prime}(y)}{P(y)}+\frac{b_{1}}{y}+\frac{b_{2}}{y-1}+\frac{b_{3}}{y+1}+c_{1}y+c_{2}\right]^{2}
+\left[
\frac{P^{{\prime}{\prime}}(y)}{P(y)}-\left(\frac{P^{\prime}(y)}{P(y)}\right)^{2}
\right] \]
\[
-\frac{b_{1}}{y^{2}}-\frac{b_{2}}{(y-1)^{2}}-\frac{b_{3}}{(y+1)^{2}}+c_{1}
 +\frac{3}{4}\frac{y^{2}}{(y^{2}-1)^{2}}-\frac{1}{2}\frac{1}{y^{2}-1}
\]
\begin{equation}
+ \frac{1}{y^{2}-1}[E+
\frac{A}{y^{2}}-\frac{B}{y^{2}-1}+Cy^{2}-Dy^{4}]=0 \label{5.2b}
\end{equation}
and  for large $ y $ equating the coefficients of $ y^{2}$ to zero
gives,
\[ c_{1}^{2}-D=0, \]
hence
\begin{equation}
 c_{1}= \pm \sqrt{D}= \pm q_{1}.
\end{equation}
For large $ y $ equating the coefficients of $ y $  gives
\[ 2c_{1}c_{2}=0,  \]
hence
\begin{equation}
c_{2}=0.
\end{equation}
The correct sign of  $c_{1} $ \ is fixed by square integrability
and is found to be  $c_{1}=-q_{1}$.   As the potential is
symmetric, there are moving poles on either side and hence we take
$P(y)$ to have the form given below
\begin{equation}
P(y)=\prod_{k=1}^{n}(y^{2}-y_{k}^{2}). \label{5.2c}
\end{equation}
The wave-function for this model is computed as follows
\begin{equation}
 \psi(y) = \exp{\int{[\chi
-\frac{1}{2}\frac{y}{y^{2}-1}]}dy}.
\end{equation}
\begin{equation}
 = \exp{\int{[\frac{P^{\prime}(y)}{P(y)}+\frac{b_{1}}{y}+\frac{b_{1}^{\prime}}{y-1}
+\frac{b_{1}^{{\prime}{\prime}}}{y+1}+c_{1}y]}dy}.
\end{equation}
On integrating and substituting the values of $ b_{1},b_{2} $ and
$c_{1}$ we get the expression for the wave-function in the $x$
variable as
\begin{equation}
\psi(x)=(\cosh ^{2}x)^{s_{1}-\frac{1}{4}}(\sinh
^{2}x)^{s_{2}-\frac{1}{4}} \exp\left({-\frac{q_{1}}{2}\cosh
^{2}x}\right)\prod_{k=1}^{n}(\cosh ^{2}x -y_{k}^{2}). \label{5.2d}
\end{equation}

{\it{\bf{Computation of energy-eigenvalue}}}:  We shall now show
how our analysis leads to the correct answer for energy spectrum.
With $ c_{2}=0$  and substituting the values of $b_{1},b_{2} $and
$c_{1}$  (\ref{5.2b}) takes the form

\[
\frac{P^{{\prime}{\prime}}(y)}{P(y)}
+\frac{P^{\prime}(y)}{P(y)}\left[\frac{(4s_{1}-1)}{y}+\frac{2s_{2}}{y-1}
+\frac{2s_{2}}{y+1}-2q_{1}y\right] \]
\[
+[q_{1}^{2}y^{2}+
\frac{4s_{1}s_{2}}{y-1}-\frac{4s_{1}s_{2}}{y+1}+\frac{s_{2}}{y+1}-\frac{s_{2}}{y-1}
+\frac{s_{2}^{2}}{y-1}-\frac{s_{2}^{2}}{y+1}-2s_{2}q_{1}\frac{y}{y+1}-2s_{2}q_{1}\frac{y}{y-1}
\]
\begin{equation}
-4s_{1}q_{1}-\frac{1}{8}\frac{1}{y^{2}-1}+
\frac{E+A}{y^{2}-1}+\frac{B}{2}\frac{1}{y^{2}-1}
+C\frac{y^{2}}{y^{2}-1}-D\frac{y^{4}}{y^{2}-1}]=0  \label{5.2e}
\end{equation}
Using (\ref{5.2c}) in (\ref{5.2e}) we get the following equation.
\[
{\left[\sum_{k=1}^{n} \frac{2y}{y^{2}-y_{k}^{2}}\right]}^{2}\]
\[-\sum_{k=1}^{n}\left[\frac{1}{(y+y_{k})^{2}}+\frac{1}{(y-y_{k})^{2}
}\right]+\sum_{k=1}^{n}
\frac{2y}{y^{2}-y_{k}^{2}}\left[\frac{(4s_{1}-1)}{y}+\frac{2s_{2}}{y-1}+\frac{2s_{2}}{y+1}-2q_{1}y
\right]
\]
\[+[q_{1}^{2}y^{2}+
\frac{4s_{1}s_{2}}{y-1}-\frac{4s_{1}s_{2}}{y+1}+\frac{s_{2}}{y+1}-\frac{s_{2}}{y-1}
+\frac{s_{2}^{2}}{y-1}-\frac{s_{2}^{2}}{y+1}
-2s_{2}q_{1}\frac{y}{y+1}-2s_{2}q_{1}\frac{y}{y-1}
\]
\begin{equation}
-4s_{1}q_{1}-\frac{1}{8}\frac{1}{y^{2}-1}+
\frac{E+A}{y^{2}-1}+\frac{B}{2}\frac{1}{y^{2}-1}+C\frac{y^{2}}{y^{2}-1}-D\frac{y^{4}}{y^{2}-1}]=0
\label{5.2f}
\end{equation}
Multiplying the above throughout by $\frac{1}{y}$ and integrating
along a closed contour enclosing $y=0$ we get the expression for
energy as

\begin{equation}
E=-4{\left[s_{1}+s-{2}-\frac{1}{2}\right]}^{2}-8s_{1}\left[
\frac{q_{1}}{2}+\sum_{k=1}^{n}\frac{1}{{y_{k}}^{2}} \right]
\end{equation}

Using $y_{k}^{2}=\xi_{k}$ the above equations for energy become

\begin{equation}
E=-4{\left[s_{1}+s-{2}-\frac{1}{2}\right]}^{2}-8s_{1}\left[
\frac{q_{1}}{2}+\sum_{k=1}^{n}\frac{1}{\xi_{k}} \right]
\label{5.2g}
\end{equation}
Changing to $y^{2}=\xi$ and  $y_{k}^{2}=\xi_{k}$  (\ref{5.2f})
becomes
\[
4\xi\left[\sum_{k=1}^{n}\frac{1}{\xi-\xi_{k}}\right]^{2}
-2\sum_{k=1}^{n}\left[\frac{\xi+\xi_{k}}{(\xi-\xi_{k})^{2}}\right]
+2\left[(4s_{1}-1)+4s_{2}\frac{\xi}{\xi-1}-2q_{1}\xi\right]\sum_{k=1}^{n}
\left[\frac{1}{\xi-\xi_{k}}\right]
\]
\[[q_{1}^{2}\xi+8s_{1}s_{2}\frac{1}{\xi-1}-2s_{2}\frac{1}{\xi-1}+2s_{2}^{2}\frac{1}{\xi-1}
-4s_{2}q_{1}\frac{\xi}{\xi-1}-4s_{1}q_{1}-\frac{1}{8}\frac{1}{\xi-1}
\]
\begin{equation}
+\frac{(E+A)}{\xi-1}+\frac{B}{2}\frac{1}{\xi-1}+C\frac{\xi}{\xi-1}
-D\frac{\xi^{2}}{\xi-1}] =0   \label{5.2j}
\end{equation}
and integrating  (\ref{5.2j}) over $\xi$ around a closed contour,
enclosing only one of the points  $\xi_{i} $ and repeating for
$(i=1,2,\cdots,n)$ we get the following result

\begin{equation}
\prod_{k=1}^{n} \frac{1}{\xi_{i}-\xi_{k}}-\frac{s_{1}}{\xi_{i}}
+\frac{s_{2}}{\xi_{i}-1} -\frac{q_{1}}{2} =0   \label{5.2h}
\end{equation}
\[ i=1,2,\cdots n\]
The results (\ref{5.2d}), (\ref{5.2g}) and (\ref{5.2h}) are in
agreement with those given in [6]

The general feature of the zeros of the wave-function for the
sextic oscillator are also true for the QES hyperbolic potential.
In particular, for a given potential it is correct that all the
exactly solvable  wave-functions have the same total number, (real
and complex) of zeros.  This feature is found to be correct for
all QES model studied so far including the QES periodic potentials
[12].

%XXXXXXXXXXXXXXXXXXXXXXXXXXXXXXXXXXXXXXXXXXXXXXXXXXXXXXXXXXXXXXXXXXXXXXXXXXXXX
%XXXXXXXXXXXXXXXXXXXXXXXXXXXXXXXXXXXXXXXXXXXXXXXXXXXXXXXXXXXXXXXXXXXXXXXXXXXXXX

\section{Concluding Remarks}
 Our study of bound state wave-functions in this chapter shows the
 following similarities and differences between the exactly
 solvable and QES models.

\begin{enumerate}
\item In both the models, the "QMF" turns out to be a rational
      function after a suitable change of variables.
\item In both the cases, the integer $n$ in the right hand side
      of quantization condition coincides with the number of
      moving poles.
\item For every bound state in one dimension the $k^{th}$ excited state
      wave-function have $k$-nodes on the real axis.  This
      statement is a general one and is true for all models
      including exactly solvable and QES potentials. The study in
      chapter 3 shows that QMF for exactly-solvable models has
      moving poles which are in correspondence with the nodes of
      the wave-function.  There are no poles off the real axis.
      However this property fails to be true for QES potentials
      where the QMF has poles off the real axis, in addition to
      the poles on the real axis corresponding to the nodes of the
      wave-function.
\item For the QES potentials only a part of the energy spectrum and
      the corresponding wave-functions can be computed exactly.
      An interesting property of the QMF for all these levels is
      that the total number of moving poles is the same and equal
      to the integer $n$ of the quantization.
\item Different values of integer $n$ correspond to different QES
      potentials within a family, and it does not refer to
      different excited state of a single potential, as was the
      case for exactly-solvable model.
\end{enumerate}

%XXXXXXXXXXXXXXXXXXXXXXXXXXXXXXXXXXXXXXXXXXXXXXXXXXXXXXXXXXXXXXXXXXXXXXX
%XXXXXXXXXXXXXXXXXXXXXXXXXXXXXXXXXXXXXXXXXXXXXXXXXXXXXXXXXXXXXXXXXXXXXXX

\chapter{CONCLUSIONS AND OUTLOOK}

In this thesis, we have studied exactly solvable and QES
potentials in one dimensional quantum mechanics with in the frame
work of QHJ formalism.  The following results have been obtained.

\begin{enumerate}
\item  The eigen-values and eigen-functions of the exactly solvable
       models can be obtained by a very simple and elegant method, which
       makes use of elementary results from theory of complex variables.

\item  The non-trivial input in this analysis is the singularity
       structure  of  the QMF.  Besides this, a change of variable
       is needed to transform the QHJ equation to a Riccati form with
       rational functions as coefficients.  Having done this, it
       is very easy to identify the fixed poles and the
       corresponding residues.  It is the location and the number
       of moving poles which present some difficulty.  For a large
       class of exactly solvable and QES models  studied by us,
       the number of moving poles for the bound states turns out
       to be finite.  In addition, the behavior  of QMF for large
       values of independent variables has been very simple to
       read from QHJ equation.  All these observations can be summarized in
       one sentence by saying that the QMF is a rational function
       after a suitable change of variables.

\item  The quantization condition as given by Leacock and Padgett,
       is applicable to separate systems which can be reduced to
       one dimensional problems.  It will be interesting  to
       formulate an exact quantization condition for
       non-separable systems in higher dimensions, and
       investigate its relation to the well known existing
       semi-classical schemes and to see applications to chaotic
       systems.

\item  For other models which are not exactly solvable, one has to
       device and approximation scheme.  Here again some idea
       about the knowledge of location of moving poles has any
       relation to classical trajectories.

\item  We have tried to study the QHJ formalism for an-harmonic
       oscillator which is a test case of any computational
       scheme.  One can compute the asymptotic value of the QMF
       for large $x$ and one can use this answer as an input for
       numerical integration of QHJ equation.  Detailed investigation is in
       progress and interesting approximation scheme for
       an-harmonic oscillator is expected from the preliminary
       results.  The simple form of QHJ  equation offers a
       possibility of several analytic approximation schemes also.

\item  When computing bound states for some potentials such as
       Rosen Morse hyperbolic potential, it is found that applying
       boundary condition $p(x,E) \stackrel{\hbar \rightarrow 0}
       \longrightarrow p_{cl}(x,E)$, carefully leads one to select
       different residues for different ranges of potential
       parameters.  Excepting this and proceeding, further
       analysis leads to different set of energy spectrum and
       wave-functions for such different ranges of parameters in
       the potential.  This is consistent with the known result on
       phases of super-symmetry in Rosen Morse potential
       [13].  Other such potentials, for example
       trigonometric Scarf potential [14], which exhibit different
       phases for different ranges of potential parameters, can
       also be investigated within our frame work.

\item  It must be remarked that, the QHJ formalism as presented in this thesis
       is applicable to bound states only.  Modifications will be
       needed to apply this formalism to continuous energy
       solutions.  The requirements such as the quantization rule,
       square integrability etc., are no longer applicable.  For
       such cases we must accept all possible combination of
       residues, consistent with the other equations of the
       theory, and proceed to analyse the consequences.  In fact
       analysis of this type has been performed for some of the QES and
       exactly solvable periodic potentials, and QHJ formalism
       leads to the full set of band-edge wave-function and
       corresponding energy eigen-values [12,15].

\item  The QHJ formalism offers advantages from the pedagogical point of
       view.  The understanding of method and results requires only
       the basic understanding of the theory of complex variables.
       In this connection we mention periodic potentials, both exactly
       solvable and QES, and PT symmetric complex potentials [16]
       which can be handled with equal ease within the QHJ approach
       [12,15, 17].

\end{enumerate}

\newpage
\noindent
\markboth{}{}

\begin{center}
{\Large REFRENCES}
\end{center}
\begin{enumerate}

\item  Leacock R A and Padgett M J, Phys. Rev. Lett. {\bf 50} (1983) 3 \\
       Leacock R A and Padgett M J, Phys. Rev. {\bf D 28} (1983) 2491
\item  Goldstein H, ``Classical Mechanics'', Addison Wesley, New York 1984
\item  Bhalla R S, Kapoor A K and Panigrahi P K, Am. J. Phys {\bf 65} (1997)
       1187
\item  Bhalla R S, Kapoor A K and Panigrahi P K, Mod. Phys. Lett. {\bf A12}
      (1997) 95

\item  Sree Ranjani S, Geojo K G, Kapoor A K and Panigrahi P K, Mod. Phys. Lett.       {\bf 19 A} (2004) 1457

\item Ushveridze A, ``Quasi-Exactly Solvable Models in Quantum
      Mechanics'' Bristol: Institute of Physics Publishing 1994

\item Geojo K G, Sree Ranjani S, and Kapoor A K, J. Phys.
      Math. Gen. {\bf A 36}  (2003) 4591

\item Cooper F, Khare A and Sukhatme U 2001 "Sypersymmetry in Quantum Mechanics (Singapore:  World Scientific) and references therein

\item Gon\'{z}alez-L\'{o}pez A, Kamran N and Olver P J, Commun.  Math. Phys {\bf 153} 117
\item Gonzale\'{z}-L\'{o}pez A, Kamran H and Olver P J, Contemp. Math {\bf 160} 113

\item Ince E L, " Ordinary Differential Equations" 1956 (Dover Publications Inc. New York)

\item Sree Ranjani S, Kapoor A K, Panigrahi P K, "A Study of QES Periodic Potentials in QHJ Formalism", quant-ph/0403196. 

\item  Bhalla R S, Kapoor A K and Panigrahi P K, Int. J. Mod. Phys.{\bf A 54}
      (1996) 951

\item Scarf F L, Phys. Rev.  {\bf 112}  (1958)  1137.

\item Sree Ranjani S, Kapoor A K, Panigrahi P K "Calculation of Band Edge Eigenfunction and Eigenvalues of Periodic Potentials through the Quantum Hamilton Jacobi Formalism", quant-ph 0312041, (to appear in Mod Phys. Lett A).

\item Bender C M, Brody D J and Jones H F, Phys. Rev. Lett.  {\bf 89} (2002) 270 and references therein.

\item Sree Ranjani S, Kapoor A K, Panigrahi P K, "A Study of PT symmetric Potentials in QHJ Formalism", quant-ph/0403054.

\item Bender Carl M , "Quasi-exactly solvable quartic potential", physics/9801007

\end{enumerate}
\end{document}